\documentclass[fleqn,usenatbib,useAMS]{mnras}

\usepackage{graphicx}	% Including figure files
\usepackage{amsmath}	% Advanced maths commands
\usepackage{multicol}        % Multi-column entries in tables
\usepackage{bm}		% Bold maths symbols, including upright Greek
\usepackage{pdflscape}	% Landscape pages
\usepackage{xcolor}
\usepackage{soul}
\usepackage[capitalise]{cleveref}
\usepackage[T1]{fontenc}
\usepackage{ae,aecompl}

\usepackage{newtxtext,newtxmath}
\title[Transfer learning for cosmological inference]{Transfer learning for multifidelity simulation-based inference in cosmology}

\author[A. A. Saoulis et al.]{Alex A. Saoulis,\thanks{Contact e-mail: \href{mailto:a.saoulis@ucl.ac.uk}{a.saoulis@ucl.ac.uk}}$^{1,2}$ %,
Davide Piras,$^{3,4}$ 
Niall Jeffrey,$^1$ 
Alessio Spurio Mancini,$^{5}$ 
Ana M. G. Ferreira,$^2$
\newauthor
Benjamin Joachimi$^1$
\\
$^1$ Department of Physics \& Astronomy, University College London, Gower Street, London, WC1E 6BT, United Kingdom \\
$^2$ Department of Earth Sciences, University College London, 5 Gower Place, London, WC1E 6BS, United Kingdom  \\
$^3$ Département de Physique Théorique, Université de Genève, 24 quai Ernest Ansermet, 1211 Genève 4, Switzerland\\
$^4$ Centre Universitaire d’Informatique, Université de Genève, 7 route de Drize, 1227 Genève, Switzerland\\
$^5$ Department of Physics, Royal Holloway, University of London, Egham Hill, Egham, TW20 0EX, United Kingdom
  }

\date{Published MNRAS 01/09/25,  https://doi.org/10.1093/mnras/staf1436}

\pubyear{{\the\year{}}}

\begin{document}
\label{firstpage}
\pagerange{\pageref{firstpage}--\pageref{lastpage}}
\maketitle

\begin{abstract}
Simulation-based inference (SBI) enables cosmological parameter estimation when closed-form likelihoods or models are unavailable. However, SBI relies on machine learning for neural compression and density estimation. This requires large training datasets which are prohibitively expensive for high-quality simulations. We overcome this limitation with multifidelity transfer learning, combining less expensive, lower-fidelity simulations with a limited number of high-fidelity simulations. We demonstrate our methodology on dark matter density maps from two separate simulation suites in the hydrodynamical CAMELS Multifield Dataset. Pre-training on dark-matter-only $N$-body simulations reduces the required number of high-fidelity hydrodynamical simulations  by a factor between $8$ and $15$, depending on the model complexity, posterior dimensionality, and performance metrics used. By leveraging cheaper simulations, our approach enables performant and accurate inference on high-fidelity models while substantially reducing computational costs.

\end{abstract}

\begin{keywords}
cosmology: cosmological parameters, large-scale structure of Universe, dark matter -- methods: statistical - software: machine learning 
\end{keywords}

%%%%%%%%%%%%%%%%%%%%%%%%%%%%%%%%%%%%%%%%%%%%%%%%%%

%%%%%%%%%%%%%%%%% BODY OF PAPER %%%%%%%%%%%%%%%%%%

% The MNRAS class isn't designed to include a table of contents, but for this document one is useful.
% I therefore have to do some kludging to make it work without masses of blank space.
\section{Introduction}

Cosmological inference is increasingly turning to machine learning (ML) techniques for improved precision, accuracy, and efficiency. In particular, simulation-based inference (SBI) has emerged as a tool to enable statistical analysis of the large-scale structure beyond traditional Gaussian likelihood-based analysis. These techniques have been applied to weak lensing measurements of cosmic shear such as the Kilo-Degree Survey \citep[KiDS,][]{von2025kids, lin2023simulation}, the Dark Energy Survey \citep[DES,][]{jeffrey2021likelihood, gatti2024dark, jeffrey2025dark}, the Subaru Hyper Suprime-Cam \citep[HSC, e.g.][]{novaes2025cosmology} and the Sloan Digital Sky Survey-III Baryon Oscillation Spectroscopic Survey \citep[SDSS-III: BOSS,][]{lemos2024field, hahn2024cosmological, thiele2024neutrino}.

Inference on shear and clustering data beyond two-point statistics has gained importance for precision cosmology, particularly as upcoming surveys prepare to probe more non-linear scales (e.g., \textit{Euclid}, \citealp{mellier2024euclid}; the Vera Rubin Observatory, \citealp{ivezic2019lsst}; the Nancy Grace Roman Space Telescope, \citealp{spergel2015wide}). A wide body of research has now explored various strategies for advancing beyond-Gaussian analysis, incorporating e.g. field-level inference, such as Bayesian Origin Reconstruction from Galaxies \citep[BORG, ][]{jasche2013bayesian, jasche2015past, jasche2019physical}, using lognormal maps \citep{xavier2016improving,leclercq2021accuracy, boruah2022map}, or higher-order statistics, such as the three-point correlation function \citep{takada2003three, schneider2003three, halder2021integrated, hahn2024cosmological}, aperture mass \citep{jarvis2004skewness, semboloni2011weak, martinet2021probing, secco2022dark}, scattering transforms \citep{cheng2020new, regaldo2024galaxy, cheng2025cosmological}, and peak counts \citep{harnois2021cosmic, zurcher2022dark}. Another important line of research, particularly for better understanding the widely discussed $S_8$ tension \citep[see e.g.][for a review]{abdalla2022cosmology} is probing the effects of systematics. For example, non-linear effects on the matter distribution become important over small scales, and become coupled with complex baryonic effects which are hard to model \citep{mccarthy2018bahamas, schneider2019quantifying, schneider2020baryonic}. Large simulation efforts have been dedicated to probing the effect of baryonic feedback at small scales \citep{mccarthy2018bahamas, villaescusa2021camels, villaescusa2022camels, schaye2023flamingo, ni2023camels, elbers2025flamingo}.

SBI uses ML models known as neural density estimators (NDEs) to model the probabilistic relationship between parameters and data empirically, allowing the method to drop the common  assumption of a Gaussian likelihood\footnote{We define SBI as using ML-based models of the likelihood (or related quantites), as distinct from using a simulation-based pipeline to estimate the Gaussian covariance of the likelihood \citep[e.g.,][]{harnois2024kids}.}. Examples of modelling choices include the posterior \citep{papamakarios2016fast, alsing2018massive, alsing2019fast, greenberg2019automatic, deistler2022truncated}, the likelihood \citep{papamakarios2019sequential, lueckmann2019likelihood}, or ratios of these quantities \citep{hermans2020likelihood, durkan2020contrastive}. This allows for various sources of error, including both measurement errors and systematic effects, to be incorporated in the likelihood by directly simulating them. SBI offers an efficiency advantage over many traditional likelihood-based techniques, enabling significant reductions in the number of simulations required to model the posterior by interpolating between them \citep{alsing2018massive, cranmer2020frontier}. In addition to this, ML-based neural compression has gained popularity for extracting summary statistics from high-dimensional observations, such as from ensembles of summary statistics, weak gravitational lensing convergence maps or dark matter density maps \citep{gupta2018non, ribli2019weak, fluri2019cosmological,  fluri2022full,  matilla2020interpreting, makinen2021lossless, jeffrey2021likelihood,lu2023cosmological, gatti2024dark, lanzieri2024optimal, lemos2024field}. ML-based compression and density estimation have received further attention for their ability to significantly improve the constraining power of the observations \citep[e.g.,][]{jeffrey2021likelihood, dai2024multiscale}. 

However, training robust and informative neural compression models is challenging, particularly when considering small datasets (for instance, fewer than $\mathcal{O}(10^4)$ data examples, see e.g. \citealp{jeffrey2025dark, bairagi2025many, park2025dimensionality}). Recent work has demonstrated that ML models fail to optimally compress low-dimensional power spectrum data in a data-limited regime \citep{bairagi2025many}. Neural compression of field-level data, on the other hand, often relies on deep learning techniques such as convolutional neural networks (CNNs), which are particularly data-hungry: for instance, \citet{jeffrey2025dark} trained a large ensemble of CNN-based neural compression models in order to mitigate against their weak performance in the absence of a large training dataset. In addition, the density modelling of SBI is also hamstrung by a lack of training data. Prior work has shown that common neural density estimation techniques underperform with limited data, yielding inaccurate and poorly calibrated posteriors \citep{lueckmann2021benchmarking,hermans2022crisis,lemos2023robust,delaunoy2024low, tucci2024eftoflss, krouglova2025multifidelity}. This makes extending SBI to more realistic cosmological models that require expensive, fine-grid hydrodynamical simulations challenging. 

Our work aims to reduce the number of expensive simulations required to perform cosmological inference by leveraging cheaper simulators. In a recent example, \cite{jia2024cosmological, jia2024simulation} use a pre-trained inference model (via neural quantile estimation, NQE) that is calibrated on a small target dataset using a quantile-shifting technique. In this study, we develop the use of transfer learning, a popular technique in the ML community that leverages data from one domain to improve performance in another \citep[see e.g.][for a survey]{zhuang2020comprehensive}. One example of transfer learning is domain adaptation, which has been used within cosmology to improve the robustness of inference with respect to uncertain physical processes; domain adaptation improves generalisation across datasets by aligning their feature representations. This can be achieved in a number of ways: introducing additional loss terms, such as maximum mean discrepancy (MMD,  \citealp{roncoli2023domain}); adversarially, by training a discriminator to minimise domain differences \citep{ganin2015unsupervised, jo2025towards,andrianomena2025towards}; or using optimal transport methods to explicitly map between latent distributions \citep{wehenkel2024addressing,andrianomena2025towards}. 

%However, it is unclear how well NQE can generalise to out-of-distribution data given the calibration quantile shifts are learned unconditionally of the observations. 

Our approach is straightforward: we perform transfer learning by first pre-training on a large corpus of cheaper, lower-fidelity data before training the model on a small set of accurate examples (a process known as fine-tuning). This widely used approach underpins foundation models, which are large, generic pre-trained models that can later be fine-tuned for specific tasks \citep{he2016deep, devlin2019bert, dosovitskiy2020image, radford2021learning, zhai2022scaling, kirillov2023segment}. Pre-training allows models to learn generalisable features, which improves performance when adapting to new, related datasets \citep{bengio2012deep, kornblith2019better, hoffmann2019machine, mishra2022task2sim, tahir2024features,Lastufka24}. Some prior work has used this approach for cosmological inference \citep{sharma2024comparative, gondhalekar2024convolutional}, but  prior to this work there had been no comprehensive investigation into whether pre-training can substantially reduce the number of accurate simulations required to perform inference. Concurrent with this work, \citet{krouglova2025multifidelity} demonstrated that the exact same principle of transfer learning is effective for standard density estimation architectures such as neural spline flows, with applications directly to SBI. Since the submission of this manuscript, two independent studies proposed enhancements to a basic transfer learning strategy, each demonstrating improved outcomes \citep{thiele2025simulation, hikida2025multilevel}.

This paper is structured as follows. Our methodology is described in \cref{sec:method}. In \cref{sec:data} we describe the multifidelity simulation suites that we use for transfer learning. \Cref{sec:model_training} presents the ML architectures and training procedures developed for this work, while the metrics used for model evaluation are introduced in \cref{sec:evaluation}. \Cref{sec:results} presents the results of our multifidelity transfer learning methodology, and compares it with a high-fidelity-only approach for two examples: \cref{sec:LH_results} explores a two-dimensional inference problem, and \cref{sec:SB28} addresses a more complicated five-dimensional inference problem, with a larger set of cosmological parameters and astrophysical nuisance parameters. Finally, we present a discussion of our results and our conclusions in \cref{sec:conclusions}.

\section{Methodology}
\label{sec:method}

\subsection{Data}
\label{sec:data}

This work introduces a simple framework for multifidelity inference on cosmological data. We utilise the CAMELS Multifield Dataset \citep[CMD,][]{villaescusa2022camels, ni2023camels}, a well-studied collection of simulations covering different fidelities and sub-grid physics models, to demonstrate our methodology. In particular, the CMD includes a gravity-only $N$-body simulation suite using \texttt{GADGET-3} \citep{springel2005cosmological}, as well as several magneto-hydrodynamical simulation suites. In this study, we use $N$-body simulations as the lower-fidelity dataset, and the IllustrisTNG CMD suites as high-fidelity simulations. These IllustrisTNG simulations were produced using the \texttt{AREPO} code \citep{springel2010pur} to solve the same sub-grid physics models as the original IllustrisTNG simulations \citep{weinberger2016simulating, pillepich2018simulating}.

\begin{figure}
\includegraphics[width=\columnwidth]{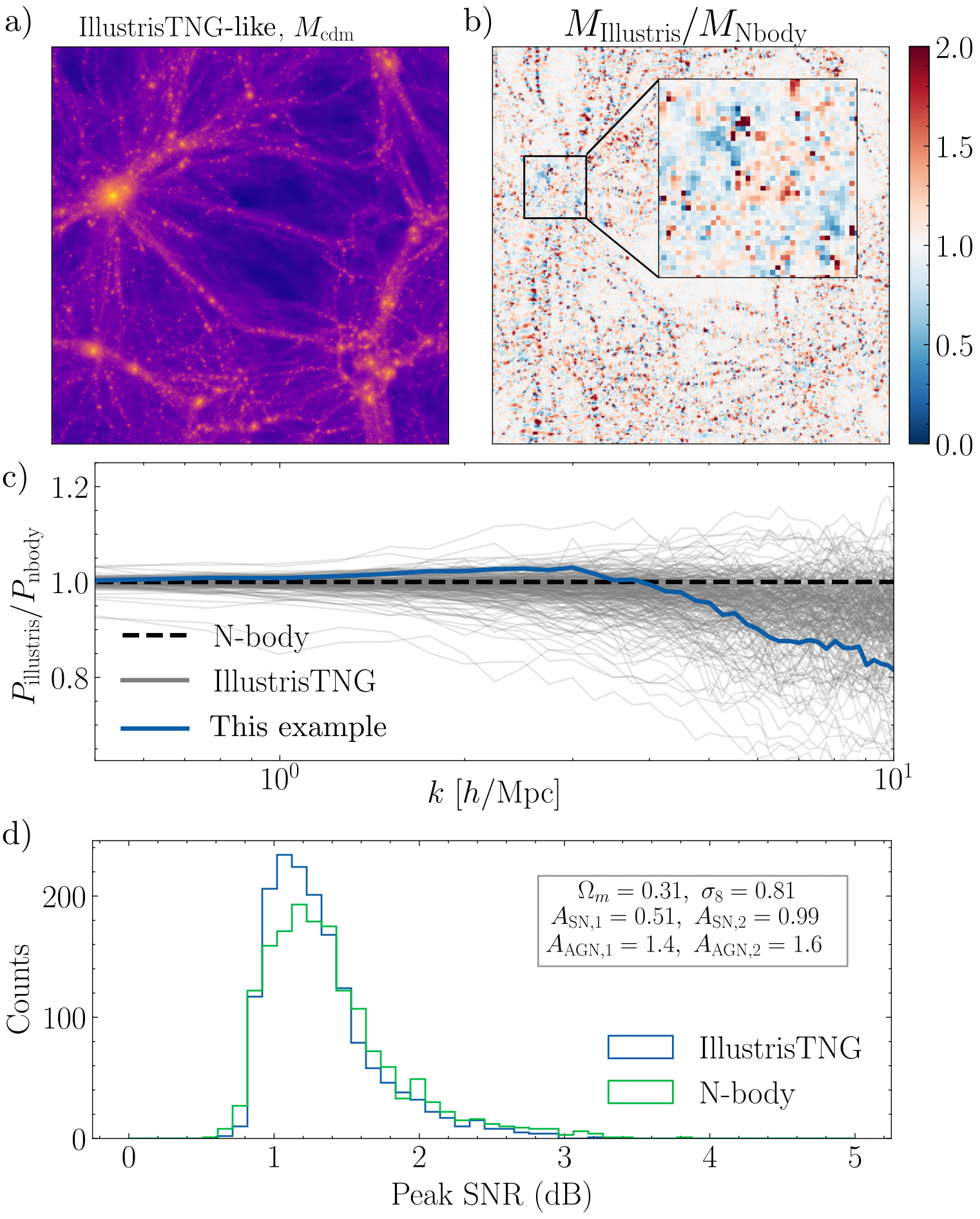}
    \caption{Comparison of the (scaled) dark matter density maps between paired simulations from the IllustrisTNG and $N$-body CMD simulation suites. Paired simulations have identical initial conditions as well as cosmological and astrophysical parameters. Panel a) shows a log-scaled dark matter density map, $M_\text{cdm}$, from the IllustrisTNG Latin hypercube (LH) suite. Panel b) gives the ratio between the IllustrisTNG $M_\text{cdm}$ and the paired $N$-body $M_\text{cdm}$, with a zoomed-in inlay of a particularly high contrast region. Panel c) shows the power spectrum ratio between the two maps (blue line), along with a random selection of 250 spectral ratios between paired maps from the LH suite (grey lines). Panel d) shows the peak count statistics, computed by applying a four pixel $M_{\textrm{ap}}$ filter (see text for details)
    and computing the peak signal-to-noise ratio (SNR) over the background level. The exact cosmological and astrophysical feedback parameters are given in panel d).}
    \label{fig:data_plot}
\end{figure}

The CMD comprises thousands of simulations sampling universes with different cosmologies and astrophysical processes. These simulations are standardised to volumes of $(25 \,h^{-1}\text{Mpc})^3$. For each simulation, 15 current-time ($z=0$) pseudo-independent 2D matter slices are extracted by considering 5 slices per dimension (of thickness $5 \, h^{-1}\text{Mpc}$); these are then pixelised into bins of approximate area $(0.1 \,h^{-1}\text{Mpc})^2$ to produce 2D images of the density fields with side of 256 pixels (see \citealp{villaescusa2021camels} for the exact procedure). This work performs inference directly on density maps: we use 2D dark matter density maps, denoted as $M_\text{cdm}$, from the $N$-body simulations and the IllustrisTNG simulations (though we demonstrate our conclusions are unchanged for total matter density maps $M_\text{tot}$ in \cref{sec:performance_ablations}). We then take the log of the maps and normalise them by subtracting the mean and dividing by the standard deviation of the pixel values, so that the resulting distribution is approximately unit Gaussian. These processed maps are then passed into the ML models. An example of these maps is given in \cref{fig:data_plot}, which shows the differences between two simulations, one hydrodynamical and one $N$-body, with identical parameters and initial conditions. \Cref{fig:data_plot} explores the differences between the multifidelity simulations through a mass density ratio map, the power spectrum ratio, and a comparison of the peak statistics. The peak statistics were computed by first applying a four pixel aperture mass ($M_{\textrm{ap}}$) filter following \citet{schneider1998new}, implemented using \texttt{lenspack}\footnote{ \href{https://github.com/CosmoStat/lenspack}{https://github.com/CosmoStat/lenspack}}.

We apply our methodology to two of the simulation suites from the CMD: (i) the Latin hypercube suite \citep[LH,][]{villaescusa2022camels}, which varies  the matter density fraction $\Omega_\textrm{m}$ and the amplitude of the matter density power spectrum, parametrised by $\sigma_8$, alongside four astrophysical nuisance parameters ($A_\textrm{SN,1}$, $A_\textrm{SN,2}$, $A_\textrm{AGN,1}$, $A_\textrm{AGN,2}$); and, (ii) the Sobol28 suite \citep[SB28,][]{ni2023camels}, which varies $\Omega_\textrm{m}$, $\sigma_8$, the scalar spectral index $n_\textrm{s}$, the Hubble parameter $h$, and the baryonic density fraction $\Omega_\textrm{b}$, in addition to $23$ astrophysical nuisance parameters. The astrophysical parameters in the LH suite control the strength and behaviour of the stellar and active galactic nuclei (AGN) feedback in the simulations, while the SB28 suite varies a more detailed set of baryonic feedback processes controlling stellar and AGN feedback, supermassive black hole growth rates, star formation rates and stellar population modelling. It is worth noting that the small simulation box-size $25 \,(h^{-1}\text{Mpc})^3$ may restrict the degree to which some of these effects, particularly AGN feedback, impact the simulations.

These suites contain 15000 and 30720 paired $N$-body and IllustrisTNG dark matter maps, respectively. In both cases we perform inference over only the cosmological parameters, implicitly marginalising over all nuisance parameters. For each suite, we reserve the last 200 cosmologies (corresponding to 3,000 matter density maps) as a holdout set, split evenly into 100 cosmologies (1,500 maps) for validation and 100 cosmologies (1,500 maps) for testing. The validation set is used for hyperparameter tuning and selection of the best model during training; the test set is used exclusively for model evaluation, with results reported in \cref{sec:results}. This strict split ensures no data leakage between the training, validation, and test sets.  The paired nature of the CMD means our transfer learning approach inherently uses seed-matched simulations at both low and high fidelity in the training set. However, we explicitly verify that we achieve the same results with unpaired low and high fidelity datasets in \cref{sec:performance_ablations}.

We augment the 2D matter map dataset by randomly flipping and rotating the images during training. We perform no further processing or augmentation, such as field smoothing or addition of noise, to the low or high fidelity maps.

\subsection{Model training}
\label{sec:model_training}

In this study, we focus on neural posterior estimation \citep[NPE,][]{papamakarios2016fast}. We perform inference directly at the map level, and train a CNN-NDE neural network end-to-end to model the posterior distribution $p(\theta \mid x)$, where $\theta$ denotes the cosmological parameters and $x$ the observation. The neural network parameters $\varphi$ are trained to produce a model of the posterior $q_\varphi(\theta \mid x)$, which is generally achieved through the forward Kullback-Leibler (KL) divergence:
\begin{equation}
    \begin{aligned}
        D_{\text{KL}}\Bigl(p(\theta \mid x) \parallel q_\varphi(\theta \mid x)\Bigr) &= \mathbb{E}_{p(\theta , x)} \left[ \log p(\theta \mid x) - \log q_\varphi(\theta \mid x) \right] \\
        &= \mathbb{E}_{p(\theta , x)} \left[ -\log q_\varphi(\theta \mid x) + \text{const.} \right].
    \end{aligned}
\end{equation}
The log-posterior term $p(\theta \mid x)$ does not depend on the neural network parameters $\varphi$, and so can be ignored as a constant in the objective function:
\begin{equation}
\label{eq:objective}
    \mathcal{L}(\varphi) = -\mathbb{E}_{p(\theta, x)} \left[\log q_\varphi(\theta \mid x) \right].
\end{equation}
Note that this formulation yields an identical objective to the Variational Mutual Information Maximisation (VMIM) approach, without the emphasis on learning an information-optimal summary statistic \citep{jeffrey2021likelihood}. 

The objective in \cref{eq:objective} is used to first pre-train the network on $N$-body dark matter maps until convergence. It is then used again without modification to fine-tune the network on IllustrisTNG maps. All weights from the pre-trained network are transferred and used to initialise the fine-tuning stage on the high-fidelity maps, with the entire network being trained during this phase. We do not freeze any layers or restrict fine-tuning to specific components (e.g. the compression network or NDE), nor do we add extra layers.  We use this framework for simplicity, since only one network needs to be trained and all training can be done end-to-end. Extensions to e.g. neural likelihood estimation, and further variants, will be explored in future work. We do not envision these would require any significant modifications, but would need several training stages as in \citet{jeffrey2021likelihood}. 

We performed an initial exploration and hyperparameter tuning of architectures for data compression and density estimation. All model architecture and hyperparameter tuning was performed on the high-fidelity-only task. This ensures that our results are not biased toward improving transfer learning performance. We were partly motivated by the fact that common neural network architectures used for neural summarisation, such as CNNs, are well-suited for transfer between datasets. This stems from the inductive bias of CNNs, which encourages the learning of generic, transferable features that are then composed over a hierarchy of scales \citep{ girshick2014rich, yosinski2014transferable, kornblith2019better}. We found that the tailored CNN architecture from \citet{villaescusa2022camels}, which was optimised for the CAMELS dataset using the hyperparameter optimisation framework \texttt{optuna} \citep{akiba2019optuna}, outperformed various standard architectures, such as ResNet \citep{he2016deep} and ConvNext \citep{liu2022convnet}, both when pre-trained from natural image data or randomly initialised. We therefore used the CNN architecture from \citet{villaescusa2022camels} as our neural compression backbone, changing only the dimension of the final output layer to instead serve as a latent embedding. We found that using the CNN to compress matter density maps to larger latent dimension sizes slightly improved performance, so set the latent dimension to $128$. 

\begin{figure}
\includegraphics[width=\columnwidth]{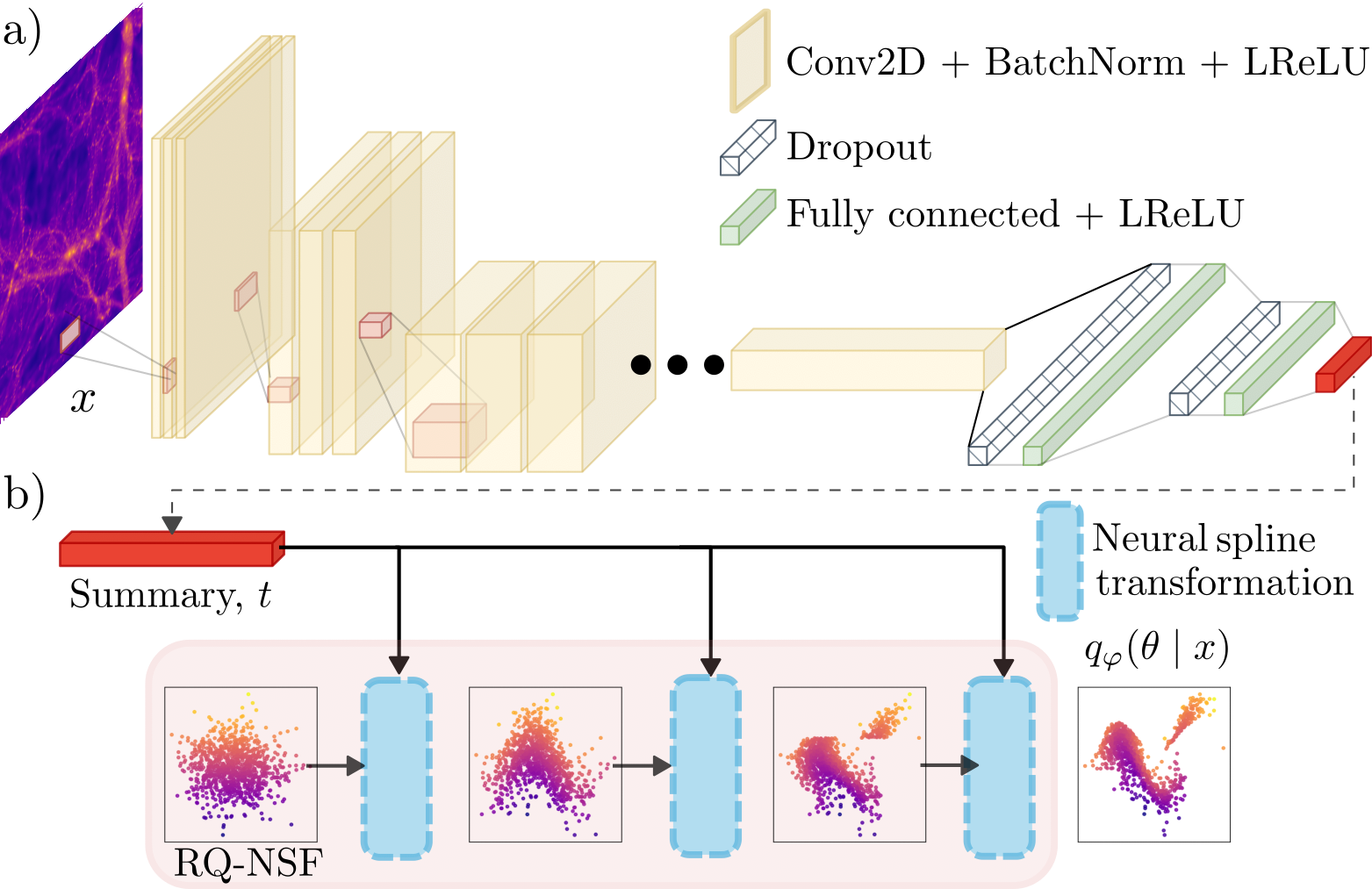}
    \caption{The neural network architecture used to perform NPE in this work. a) A CNN identical to that used in \citet{villaescusa2022camels} extracts informative features from the input dark matter map images. The CNN is built of blocks with three repeats of a 2D convolution, batch normalisation, followed by leaky ReLU non-linearities \citep[LReLU,][]{maas2013rectifier}. After each block, the spatial dimension of the activation maps is halved using a stride of $2$, and the number of channels is doubled. The final convolutional layer is flattened and passed to a pair of feedforward layers. b) The extracted low-dimensional summary statistics, $t$, are fed into a rational-quadratic neural spline flow (RQ-NSF). The RQ-NSF uses the summary statistic as conditional information to transform a simple base distribution into the modelled posterior $q_\varphi(\theta \mid x)$.}
    \label{fig:architecture}
\end{figure}

We used a rational-quadratic neural spline flow \citep[RQ-NSF,][]{durkan2019neural} as the NDE head of the network, implemented using the \texttt{sbi}  \texttt{Python} package \citep{tejerosbi2022code}. We found this NDE architecture gave significant improvements over alternative popular choices, such as Masked Autoregressive Flows \citep[MAFs,][]{papamakarios2017masked}. We also found that inserting batch normalisation \citep{ioffe2015batch, santurkar2018does} layers between spline flow blocks (similar to the CNN architecture) substantially improved performance, both in pre-training and fine-tuning. An overview of the architecture is presented in \cref{fig:architecture}.

Once the architecture was selected, we performed end-to-end training of the network on the NPE objective in \cref{eq:objective}. We made a range of modifications that improved performance. We utilised a weight decay of $0.01$, which regularises the neural network by adding a small penalty term to the network weight magnitudes \citep{krogh1991simple,loshchilov2017decoupled}. This was particularly important for the fine-tuning stage, where regularisation while training on very small datasets was greatly beneficial. We used a short learning rate (LR) warm-up period, which has been found to improve deep learning model training \citep{he2016deep, goyal2017accurate, vaswani2017attention} with a number of posited explanations \citep[see e.g.,][]{gotmare2018a,kalra2024warmup}. Larger batch-sizes (we selected 64) also improved performance for all models. We found that using a cyclic learning rate scheduler \citep{smith2017cyclical} as in \cite{villaescusa2022camels} improved performance when training on large datasets, i.e. $> \mathcal{O}(10^4)$ maps. Baseline experiments and pre-training were performed with a LR of $2\times10^{-4}$. Fine-tuning was performed with a LR of $1 \times 10^{-5}$ and an exponential decay scheduler. All models were trained using the AdamW optimizer \citep{loshchilov2017decoupled}. 

We repeat all training runs six times, changing only the initialisation of the network, the random selection of $N$ maps from the training dataset, and the (random) order in which the training data is passed to the network. The model with the lowest validation loss from each run is saved and used for evaluation. All models pre-trained on $N$-body simulations used the entire $N$-body training suite, and models with the lowest validation loss were saved for fine-tuning and evaluation.

\subsection{Evaluation}
\label{sec:evaluation}

\subsubsection{Posterior Accuracy}

Model evaluation in SBI is a well-studied task, and commonly used metrics include posterior-predictive checks \citep[e.g.,] []{papamakarios2017masked, durkan2019neural}, kernel-based distance tests such as MMD, and classifier 2-sample tests \citep{friedman2004multivariate, lopez2017revisiting}. \citet{lueckmann2021benchmarking} presented a review and comparison between various evaluation choices. In our case, where the true posterior is unknown, one could construct a ``high-quality'' posterior by using the full simulation suite to train multiple models with different initialisations, and then averaging their inference results. This ``ensemble''-based strategy can be used to improve posterior quality by integrating over the model's epistemic uncertainty \citep{hermans2022crisis, lin2023simulation}. Unfortunately, we found empirically that five-member model ensembles led to under-confident (conservative) posteriors. This finding is compatible with prior work \citep{hermans2022crisis}. We expect that this could be overcome with larger ensembles, but this approach becomes highly computationally demanding for deep learning-based models. We therefore avoid overreliance on metrics that require reference posteriors. 

In the absence of a high-quality reference posterior, the most appealing headline metric to quantify model performance is simply the mean test posterior probability (MTPP) $\log q_\varphi(\theta \mid x)$ at the true parameter values $\theta$. This is estimated by computing an expectation over the test dataset $\mathcal{D}$: 
\begin{equation}
\text{MTPP} = \mathbb{E}_{(x, \theta) \sim \mathcal{D}} \left[\log q_\varphi(\theta \mid x)\right].
\end{equation}
This serves as a robust test for posterior quality given enough test examples \citep{lueckmann2021benchmarking}, though the scale of this metric is not particularly interpretable. 

To complement this, we estimate the calibration of each model over the entire test set. This is achieved by running inference on each test data realisation, and estimating the frequency at which the truth lies within a given credibility level. This test allows us to identify modelling issues in the posteriors, such as bias and overconfidence \citep{hermans2022crisis}. For $K$ credibility level bins, we estimate the observed frequency within a given bin $ \hat{p}_i $, and compare it with the expected (ideal) frequency $p_i=1/K$. We then compute a calibration error $\mathcal{C}$, which is the relative mean squared error between the two quantities:
\begin{equation}
\label{eq:calibration_error}
    \mathcal{C} = \frac{1}{K}\sum_{i=1}^{K} \left(\frac{\hat{p}_i - p_i}{p_i}  \right)^2  = \frac{1}{K}\sum_{i=1}^{K} \left(\frac{\hat{p}_i}{p_i} -1  \right)^2.
\end{equation}
The right-hand side simply provides an equivalent expression for the calibration error $\mathcal{C}$ in terms of the ratios between observed and expected frequencies at given credibility levels, $\hat{p}_i /p_i$, which we refer to as ``overcoverage.''

We utilise Tests of Accuracy with Random Points \citep[TARP,][]{lemos2023sampling} to estimate the coverage statistics efficiently. We bootstrap the estimated credibility level statistics produced by TARP 25 times and quote the mean of the estimated $\hat{p}_i$.

 \begin{figure*}
    \includegraphics[width=\textwidth]{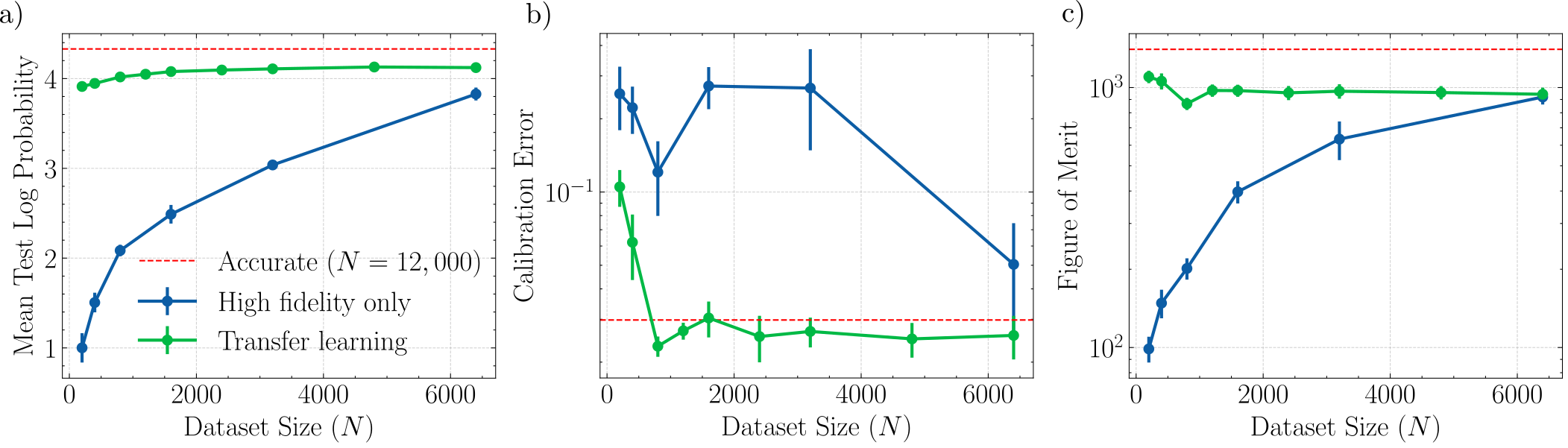}
    \caption{Inference results on the IllustrisTNG LH suite for a 2D posterior over $\Omega_\textrm{m}$ and $\sigma_8$. We compare the performance of the two approaches: training with only high-fidelity maps (blue) against the transfer learning approach (green), which uses $12000$ $N$-body maps for pre-training. An accurate benchmark model trained on the entire IllustrisTNG LH suite training set is shown by the dashed red line. The accurate model (red) uses a cyclic learning rate scheduler, which performs slightly better than an exponential scheduler for large datasets, as detailed in \cref{sec:model_training}. Panel a) shows the mean test posterior probability (MTPP), panel b) shows the calibration error $\mathcal{C}$ of the modelled posterior, defined in \cref{eq:calibration_error}, and panel c) shows the FoM, all as a function of high-fidelity dataset size $N$. Panel c) should be interpreted with the proviso that the model must be well-calibrated before the FoM measures genuine constraining power. All results show the mean and standard error over six independent training runs (including independent pre-training runs on the $N$-body simulations).}
    \label{fig:LH_metrics}
    
\end{figure*}

Together, the MTPP and calibration error $\mathcal{C}$ provide robust diagnostics of the fidelity of the learned posteriors. Once these diagnostics indicate well-calibrated and accurate inference, we can begin to assess how informative the posteriors are.

\subsubsection{Constraining Power}

Good calibration is a necessary pre-condition for a useful, reliable posterior model. Once this property is satisfied we can test the amount of information that the model is capable of extracting from the dark matter density maps. We compute the Figure of Merit (FoM), which estimates the constraining power of each model. Assuming a flat prior on $[0,1]$, the FoM can be computed as:
\begin{equation}
\text{FoM} = \left[ \det \operatorname{Cov}[\theta \mid x] \right]^{-1/n},
\end{equation}
where \(\operatorname{Cov}[\theta \mid x]\) is the covariance matrix of the posterior, and $n$ is the dimensionality of the posterior. A higher FoM indicates a tighter constraint on the parameters, meaning the model is more informative (provided the model is unbiased). In practice, we transform the cosmological variables to a flat prior on $[0,1]$ to compute the FoM and estimate the covariance using samples from the modelled posterior.

Finally, for a more interpretable metric of the posterior quality, we compute the mean squared error (MSE) between the modelled posterior mean $\hat{\theta}$ and the true parameters  $\theta$ over the entire test set. We report these MSEs broken down by parameter to probe whether the posterior quality differs significantly between parameters.

\section{Results}
\label{sec:results}

We run a range of experiments comparing  high-fidelity-only models against the transfer learning approach. We use $N$ to denote the number of IllustrisTNG 2D dark matter density maps used during the training stage. Of the dataset size $N$, $90 \%$ is used as training data, while $10\%$ is used for validation data. All results report the mean performance on the test set. 

\subsection{CAMELS Multifield Dataset: LH suite}
\label{sec:LH_results}

We compare the performance of each method over a range of IllustrisTNG dataset sizes. We present the results of the LH experiments in \cref{fig:LH_metrics}. We find that small IllustrisTNG dataset sizes lead to poor performance when training from random network initialisation, whereas pre-training on $N$-body maps leads to good performance with very few simulations. For instance, a pre-trained model that is then fine-tuned with $N=200$ IllustrisTNG maps has higher MTPP than the high-fidelity-only approach with $N=6400$ maps.

\begin{figure*}

	\includegraphics[width=0.95\textwidth]{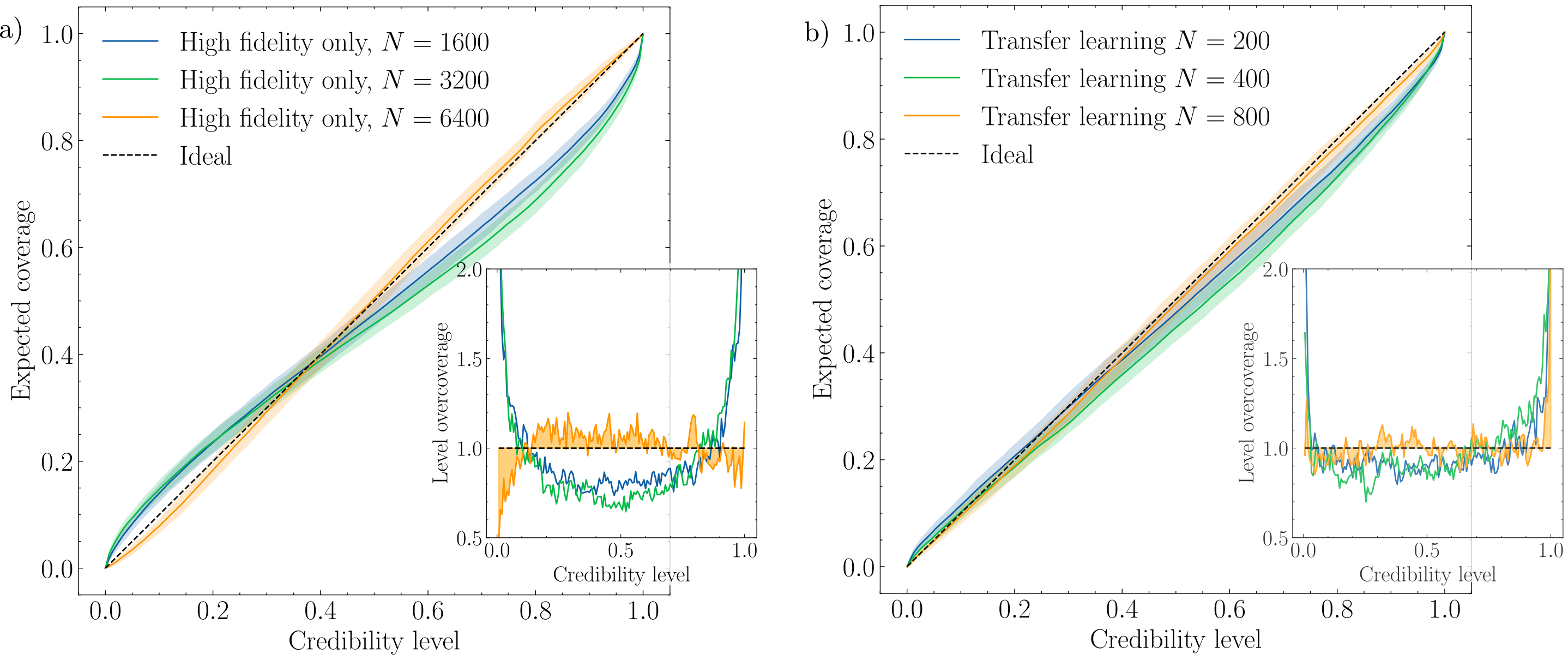}
    \caption{\textit{Main panels}: cumulative calibration curves of the nominal credibility level distribution, assessing the posterior coverage quality as a function of dataset size. The ideal calibration curve is shown by the black dashed line. Shaded regions show the $2 \sigma$ uncertainties derived from bootstrapping. \\ \textit{Insets}: the overcoverage values per credibility level, see \cref{eq:calibration_error} and the surrounding text for details. The shaded orange region highlights the discrepancy between the ideal and observed distribution of credibility levels, which is quantified by the calibration error metric introduced in \cref{eq:calibration_error}. Panel a) shows training from high-fidelity-only simulations, where models with small dataset sizes display significant overconfidence, and better calibration (though now mildly underconfident) is achieved at $N=6400$. Panel b) shows that transfer learning requires around $N=800$ IllustrisTNG maps to achieve good calibration.}
\label{fig:calibration_curves}

\end{figure*}

\begin{figure*}

	\includegraphics[width=\textwidth]{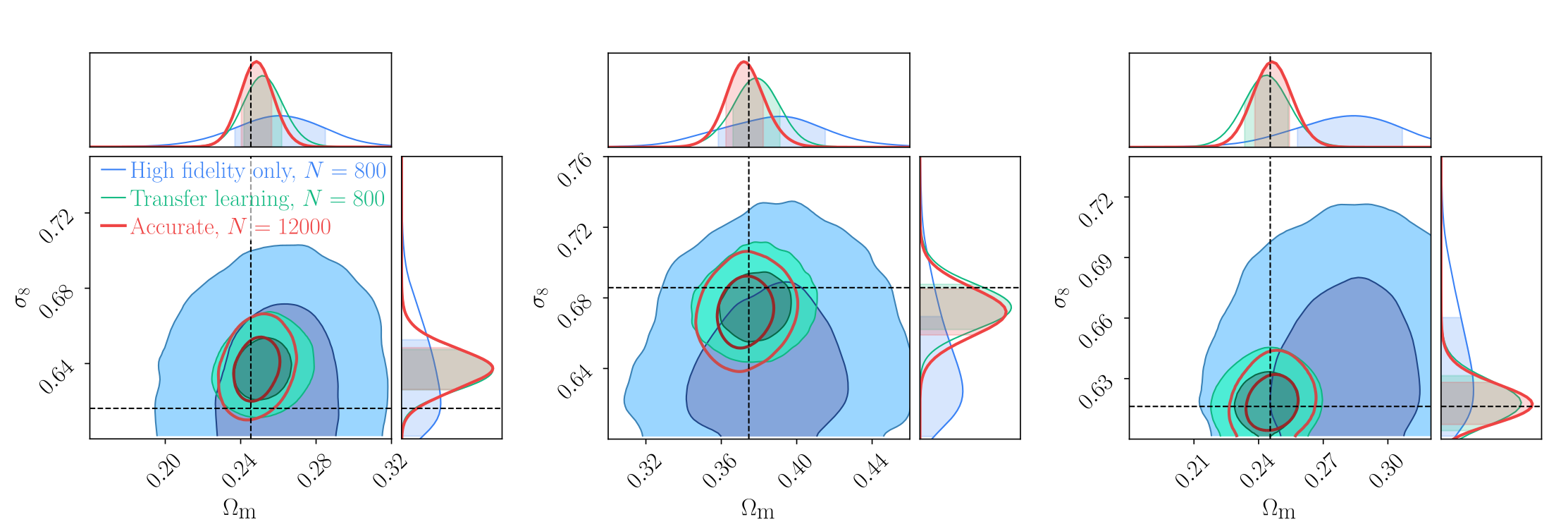}
    \caption{Three representative examples of inference from the LH simulation suite. The true cosmology is shown by the black dashed line. A model trained using transfer learning with $N=800$ high-fidelity IllustrisTNG maps is compared against a high-fidelity-only model trained with $N=800$ maps. The posteriors are compared with an ``accurate'' posterior model that was trained using the full simulation suite.}
    \label{fig:example_figure}
    \label{fig:LH_inference_posteriors}
    
\end{figure*}

One complicating factor is the posterior calibration, quantified in \cref{fig:LH_metrics}b. This demonstrates that despite the high test posterior probability, models fine-tuned with very few IllustrisTNG maps appear to be poorly calibrated. We find that acceptable calibration is achieved after $N=800$ fine-tuning maps, while training with only high-fidelity maps requires $N=6400$ for similar posterior calibration. We therefore find at least a factor of $8$ reduction in the number of simulations required to produce a performant, well-calibrated model of the posterior. 

We present calibration curves from a range of dataset sizes across the two approaches in \cref{fig:calibration_curves}. These show the standard cumulative distribution of observed credibility levels in the main panels, as well as the overcoverage distribution $\hat{p}_i/p_i$ in the insets. These reaffirm the calibration issues identified in \cref{fig:LH_metrics}: training from scratch with fewer than $N=6400$ maps leads to significantly overconfident posteriors. The high overcoverage at $\{0,1\}$ (paired with the below-ideal coverage in the middle of the distribution) is a clear indicator of overconfidence, since it indicates that the true parameters occur at extreme credibility levels too often.  On the other hand, the transfer learning models display a more minor form of bias and overconfidence until reaching around $N=800$ maps.

The FoM performance as a function of dataset size is shown in \cref{fig:LH_metrics}c. When training using only high-fidelity simulations, low dataset sizes lead to low FoMs. This observation indicates that the features (or summary statistics) extracted by the CNN are not particularly informative. Additionally, the overconfident posteriors indicate limitations in the performance of the NDE, since it is incapable of producing trustworthy posteriors. We therefore conclude that both the CNN neural compressor and the density estimation model perform poorly with small training datasets.

On the other hand, the very high MTPP and FoMs of the transfer learning models (even from $N=200$) indicate that the pre-trained CNNs produce highly informative features. However, the inferred posteriors for $N=[200,400]$ are biased and overconfident, suggesting that the NDE needs at least $N=800$ maps to correctly adjust the inferred posteriors to ensure good calibration. We found that the FoM for the $N$-body pre-training task was $\sim1400$,  much greater than the baseline and transfer learning models on the IllustrisTNG inference task. The dip in the FoM at $N=800$ is thus potentially related to the reduced constraining power of extracted CNN features on the IllustrisTNG data finally becoming properly incorporated by the NDE. This could be due to the feature-shift between $N$-body simulations and IllustrisTNG, as well as the inherent greater uncertainties due to the more complex physics of the hydrodynamical simulation suite. These results indicate that for very small fine-tuning datasets, the performance bottleneck on this task is adapting the NDE. In future work we will explore whether the NDE head could be fine-tuned while preserving good calibration statistics with more advanced techniques \citep[such as balanced SBI, e.g.][]{delaunoy2022towards}. 

Three representative examples of inference on test cosmology maps are shown in \cref{fig:LH_inference_posteriors}. We compare models produced using only high-fidelity maps ($N=800$), transfer learning ($N=800$), and an ``accurate'' reference model trained on the full IllustrisTNG training set. These examples are qualitatively consistent with the analysis presented above. The high-fidelity-only $N=800$ model gives very uninformative constraints compared to the other two posteriors. On the other hand, the fine-tuned model appears well-calibrated, and only slightly less constraining than the model trained with $\times 16.25$ more high-fidelity maps. We present a similar comparison with a high-fidelity model trained on $N=3200$ in \cref{app:further_comparison}, demonstrating that even for larger high-fidelity dataset sizes, pre-training yields significantly improved posteriors.

\cref{sec:performance_ablations} presents a range of further tests into the model performance. We found that the pre-trained models performed very poorly on high-fidelity maps when no fine-tuning was performed (corresponding to $N=0$). We explored the quality of the fine-tuned CNN compressor by freezing the CNN and re-training the NDE with larger dataset sizes. This presented more evidence that the limiting factor at very low fine-tuning dataset sizes was the NDE. We also showed that the small performance gap between the ``accurate'' model and the transfer learning models with larger dataset sizes was caused by slightly worse compression. 

Another consideration was the possibility that the paired aspect of the datasets was responsible for the significant performance gains from lower-fidelity pre-training. While we did make any explicit use of the simulation pairs, this could still have had an impact depending on the training dynamics of the network. We tested this in \cref{sec:performance_ablations} and found strong evidence that the paired aspect of the source and target dataset has no impact on our results. We therefore conclude that our transfer learning approach does not depend on paired datasets. In principle, this means that large, pre-existing simulation suites could be used as multifidelity datasets without the need to pair initial conditions and cosmological parameters. 

We found that fine-tuning with the entire IllustrisTNG suite produced a slightly weaker model than the ``accurate'' benchmark model (i.e. the transfer learning curve in \cref{fig:LH_metrics}a does not intersect with the ``accurate'' performance). We analyse the convergence properties of the approaches further in \cref{sec:performance_ablations}. We show that the slightly lower transfer learning MTPP plateau is partially due to the low learning rate (LR) used during fine-tuning, and a higher LR nearly recovers the ``accurate'' performance. This is consistent with the intuitive notion that a higher LR allows the training procedure to escape from the slightly sub-optimal region of the weight-space that is reached during pre-training. These observations are perfectly compatible with the study of \citet{sharma2024comparative}, who found no clear benefits of transfer learning when using a large high-fidelity dataset.

\begin{figure}

	\includegraphics[width=\columnwidth]{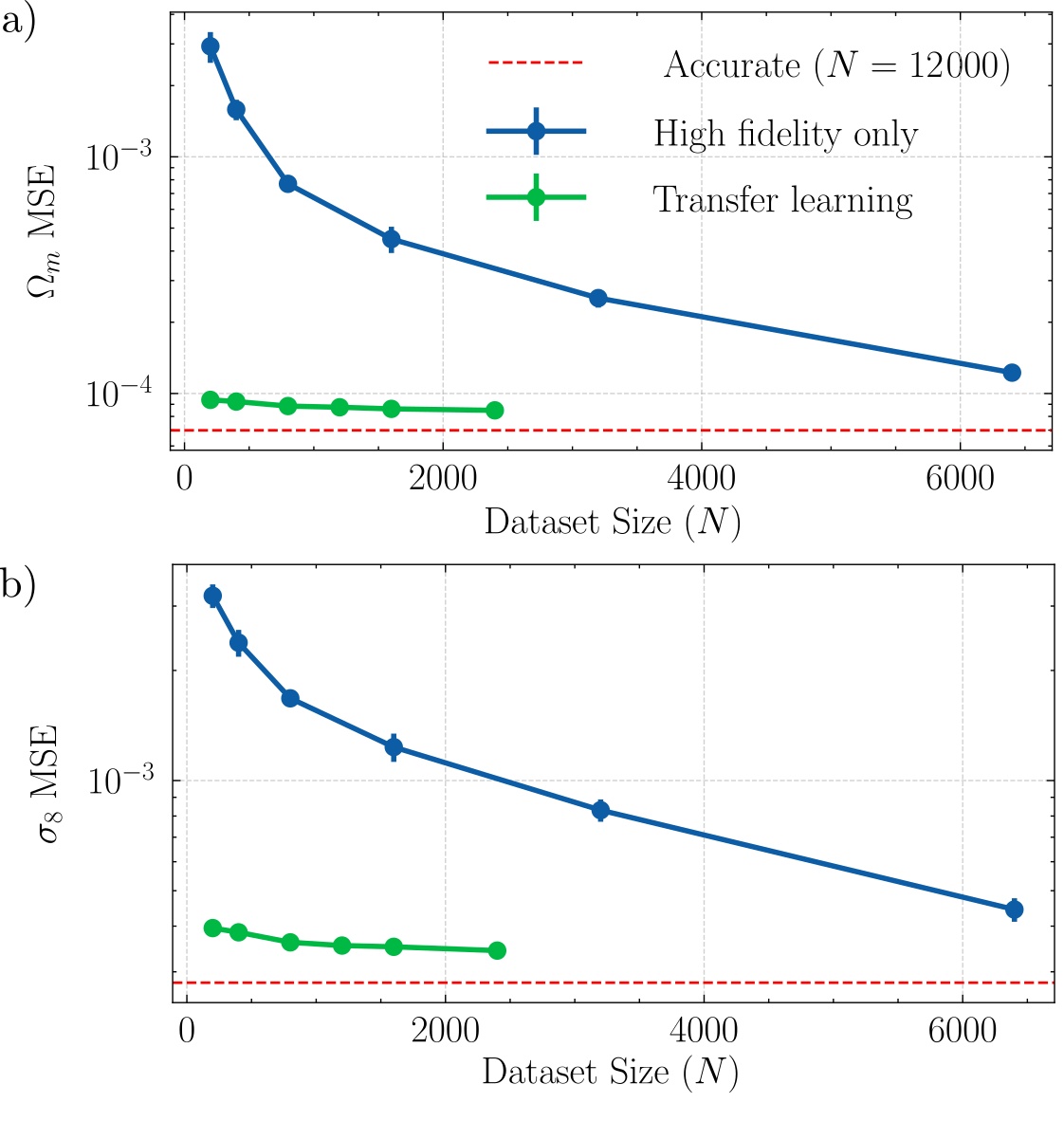}
    \caption{Mean squared error (MSE) between the inferred posterior mean $\hat{\theta}$ and the true cosmology $\theta$. Results are broken down per-parameter, with panel a) showing $\Omega_\textrm{m}$ and panel b) showing $\sigma_8$. Transfer learning models show very good consistency with the true cosmological parameters. }
\label{fig:LH_parameter_errors}

\end{figure}

We present the degree of agreement between the posterior sample means $\hat{\theta}$ and the true cosmologies $\theta$ in \cref{fig:LH_parameter_errors}, broken down for $\sigma_8$ and $\Omega_\textrm{m}$. These indicate pre-training yields very large improvements in the posterior for both parameters. $\Omega_\textrm{m}$ and $\sigma_8$ are inferred with similar precision relative to the baseline over the entire test set.

\begin{figure*}

	\includegraphics[width=\textwidth]{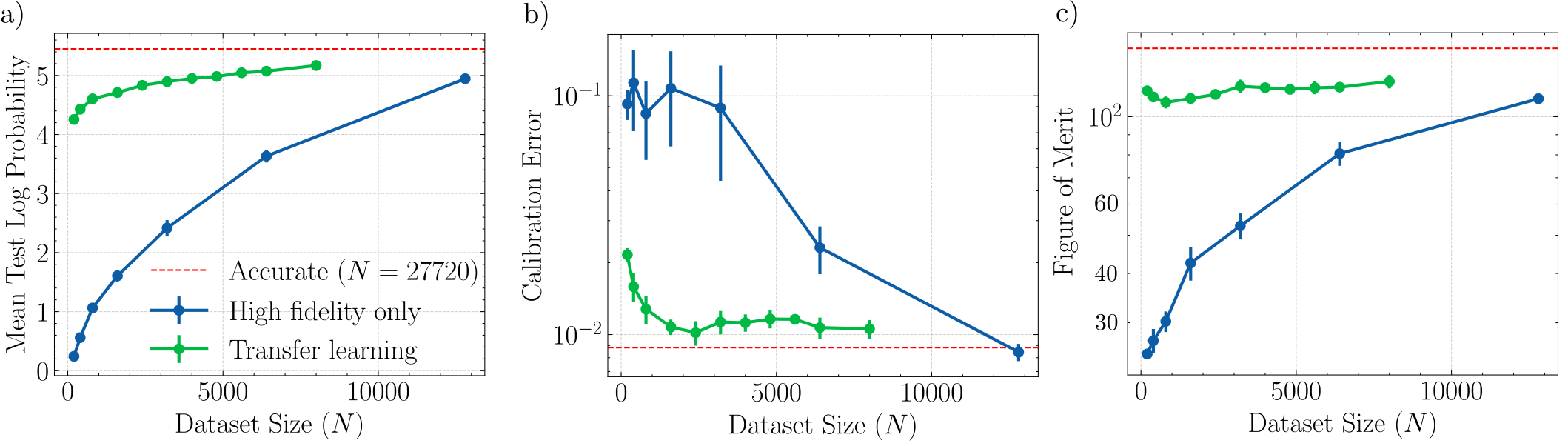}
    \caption{Inference results for the 5-dimensional posteriors on the IllustrisTNG SB28 suite. Training with only high-fidelity maps (blue) is compared against the transfer learning approach (green). An accurate benchmark model trained on the entire IllustrisTNG SB28 suite training set is shown by the dashed red line. Panel a) shows the MTPP, panel b) shows the calibration error and panel c) shows the FoM, all as a function of high-fidelity dataset size $N$. Panel c) should be interpreted with the proviso that the model must be well-calibrated before the FoM measures genuine constraining power.  Again, all results show the mean and standard error over six independent training runs.}
    \label{fig:SB28_metrics}
\end{figure*}

\begin{figure*}

	\includegraphics[width=\textwidth]{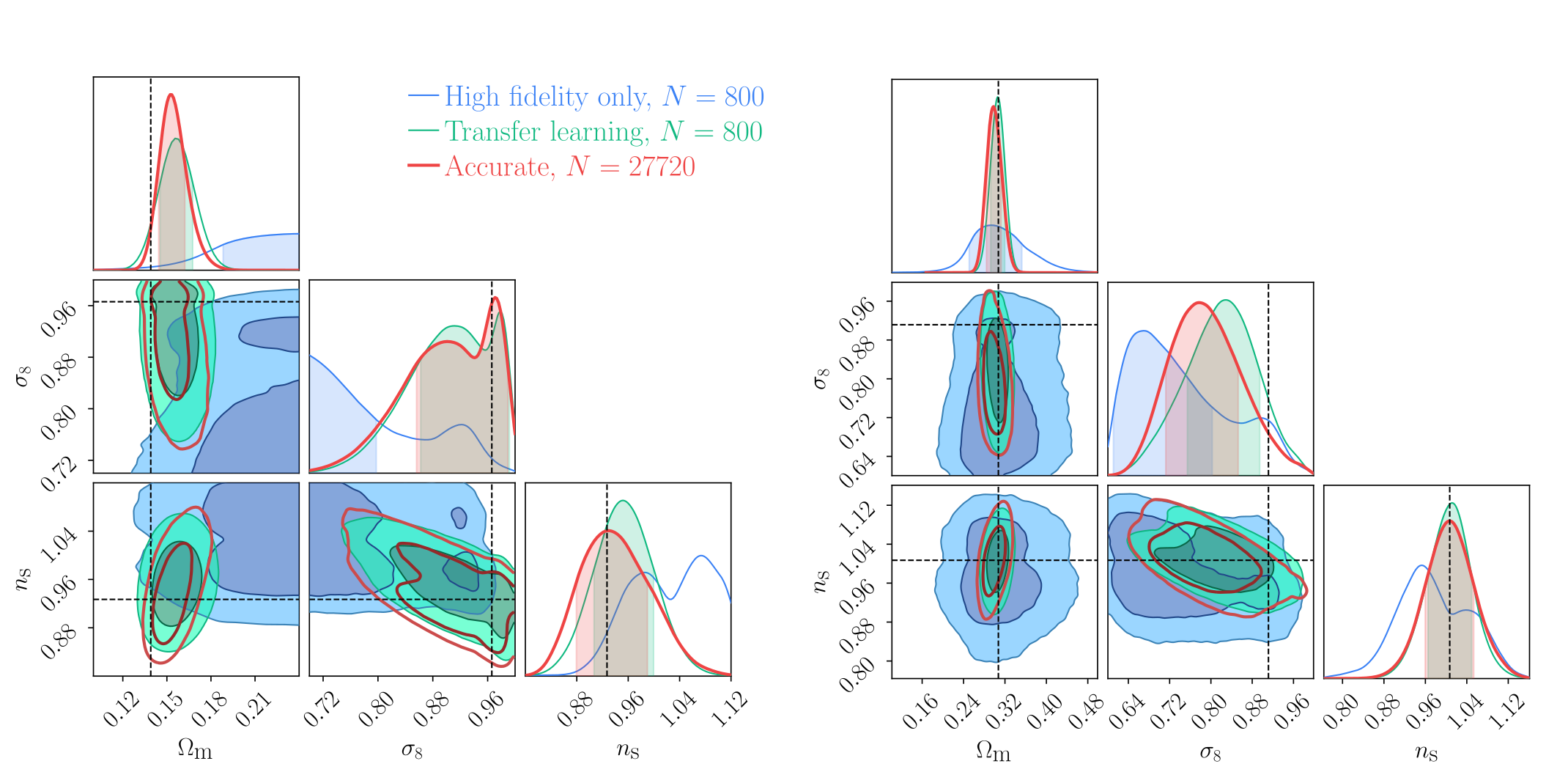}
    \caption{Two examples of posterior inference on IllustrisTNG dark matter maps from the SB28 test set. Contours are visualised over the three parameters that can be constrained by the data: $\{\Omega_\textrm{m}, \sigma_8, n_\textrm{s}\}$. The true cosmology is shown by the black dashed line. A model trained using transfer learning with $N=800$ high-fidelity IllustrisTNG maps is compared against a high-fidelity-only model trained with $N=800$ maps. The posteriors are compared with an ``accurate'' posterior model that was trained using the full training set.}
    \label{fig:SB28_posterior_examples}
\end{figure*}

\subsection{CAMELS Multifield Dataset: SB28 suite}
\label{sec:SB28}

We repeat the experiments of \cref{sec:LH_results} on the SB28 suite, this time performing inference over a 5-dimensional posterior. The very broad range of nuisance parameters, as well as the extra three cosmological parameters $\{n_\textrm{s}, h, \Omega_\textrm{b}\}$, lead to a more challenging inference problem. Prior work has only explored inferring $\Omega_\textrm{m}$ and $\sigma_8$ from this dataset, and has found that the larger set of cosmological and astrophysical parameters leads to much weaker constraints on $\sigma_8$ \citep{ni2023camels}. \citet{ni2023camels} also indicated that the Hubble constant $h$ and the baryonic fraction $\Omega_\textrm{b}$ have  minor effects on the simulations, indicating that these may be challenging to constrain. Note that the SB28 suite contains roughly double the number of simulations as the LH suite, enabling a more accurate ML-based reference model.

\cref{fig:SB28_metrics} displays the headline metrics comparing the two training approaches with a baseline ``accurate'' model that used the entire SB28 suite training set. Again, the transfer learning models significantly outperform models trained from scratch, and the MTPP score of the transfer learning experiments are only surpassed when training a high-fidelity-only model with $N=12,800$ IllustrisTNG maps. Panels b) and c) in \cref{fig:SB28_metrics} present a similar pattern as in \cref{sec:LH_results}: all transfer learning models are more constraining and better calibrated than training from scratch (until $N=12,800$). However, fine-tuned models display (minor) calibration issues until $N\ge 800$. The uptick in calibration error $\mathcal{C}$ for the transfer learning approach is very minor and largely within errors, so we do not attempt to interpret it.  Interestingly, all fine-tuned models are better calibrated in the SB28 experiments than in the LH experiments (note the different $y$-axis scales). 

The performance difference between high-fidelity-only models and transfer learning models is even larger than in \cref{sec:LH_results}. Depending on the exact MTPP performance desired, \cref{fig:SB28_metrics} indicates that pre-training on $N$-body simulations allows for a factor of $10$ to $15$ reduction in high-fidelity simulations to train an informative, well-calibrated model of the posterior. 

We found that none of the models could constrain $\Omega_\textrm{b}$ and $h$ far beyond the uniform prior. Given that this was a feature of the ``accurate'' baseline model, trained on $27720$ maps, we may conclude that this a genuine feature of the simulations, at least up to the resolving power of the CNN-NDE architecture used to perform inference. This is reflected by the FoM results in \cref{fig:SB28_metrics} c), which are significantly lower for all models than the FoM in the LH suite experiment. We present two examples of inference in \cref{fig:SB28_posterior_examples}, showing only $\{\Omega_\textrm{m}, \sigma_8, n_\textrm{s}\}$. This implicitly marginalises over the poorly constrained $\Omega_\textrm{b}$ and $h$. Again, we compare training from scratch with $N=800$ and fine-tuning with just $N=800$ maps against an ``accurate'' baseline. We present examples of inference of the full 5-dimensional posterior in \cref{app:SB28}.

Again, multifidelity transfer learning produces a model that significantly outperforms high-fidelity-only training. The difference in posterior quality is even more stark than in \cref{fig:LH_inference_posteriors}; the high-fidelity-only posteriors are very uninformative and fail to extract much useful cosmological information from the density maps. The overconfidence identified in \cref{fig:SB28_posterior_examples} is apparent as a bias in the left panel of \cref{fig:LH_inference_posteriors}. On the other hand, the transfer learning approach recovers the key features of the accurate baseline posteriors, including both the location and width of the posterior contours. Both the ``accurate'' and transfer learning models yield a degeneracy between the amplitude of the matter density power spectrum $\sigma_\textrm{8}$ and the scalar spectral index $n_\textrm{s}$. This degeneracy is expected for two-point statistics at the very short scales probed in the CMD simulations, given both parameters have similar, difficult to distinguish marginal effects on the power spectrum. The results in \cref{fig:SB28_posterior_examples} indicate that the non-linear effects probed by the CNN are insufficient to fully break this degeneracy. 

The larger performance gains reported here relative to the LH suite in \cref{sec:LH_results} likely result from the more complex task of modelling a 5-dimensional posterior and marginalising over a larger set of astrophysical parameters. We tentatively conclude that transfer learning may perform even better in more challenging inference problems, particularly those involving higher-dimensional posteriors and a broader set of nuisance parameters. As the complexity of the target task increases, the value of incorporating prior knowledge through pre-training is likely to grow.

\section{Discussion and Conclusions}
\label{sec:conclusions}

In this study, we have demonstrated that leveraging multifidelity simulations can significantly reduce the number of expensive simulations required to perform cosmological inference with SBI. By pre-training a neural inference model on a large set of lower-fidelity dark matter only simulations, we were able to perform informative and well-calibrated inference on IllustrisTNG hydrodynamical simulations with  $< 1000$ high-fidelity dark matter maps. This is a substantial improvement over previous work, which had demonstrated that training neural compression algorithms with small datasets led to suboptimal compression and inference \citep{hermans2022crisis,park2025dimensionality, bairagi2025many, jeffrey2025dark}. The relative simplicity of our framework makes this method broadly applicable across cosmology.

Prior work has explored various approximate Bayesian computation (ABC) methods for multifidelity inference \citep{prescott2020multifidelity, prescott2021multifidelity}, for instance by using low fidelity rejection sampling to improve the accurate simulation efficiency during inference. These have been extended to sequential multifidelity ABC \citep{warne2022multifidelity}, as well as to likelihood-free multifidelity inference by leveraging importance sampling \citep{prescott2024efficient}. Variance reduction strategies that capitalise on paired multifidelity simulations to isolate the statistical uncertainty, have also been used to improve estimates of cosmological observables \citep{chartier2021carpool, lee2024zooming}. Adaptation of these strategies directly for cosmological inference, particularly when dealing with significantly non-Gaussian posteriors that SBI is well suited for, remains an interesting avenue for future work. An additional line of work would be to explore tailoring our approach for transfer learning, for instance through architecture improvements that work with Fourier representations of the inputs \citep[e.g.,][]{yang2020fda, mao2023intriguing, bernardini2025ember} or specialised pre-training and transfer learning techniques \citep[e.g.,][]{he2022masked, oquab2024dinov2, akhmetzhanova2024data}. 

This work focused on matter density maps at different fidelities; there are many more possible observables that have previously been probed for performing cosmological inference, such as neutral hydrogen, gas temperature and metallicity maps \citep{hassan2020constraining, prelogovic2022machine, andrianomena2023predictive, andrianomena2025towards, gluck2024observationally}. Adaptation between observables could call for similar specialised approaches \citep[e.g.,][]{lian2025isolated}.

Future work could apply transfer learning to a wide variety of multifidelity datasets across cosmological inference. Recent work has demonstrated that neural compression even performs sub-optimally on lower dimensional data, such as power spectra \citep{bairagi2025many} or ensembles of traditional summary statistics \citep{park2025dimensionality}, when dataset sizes are limited. There is a very wide array of methods for producing mock observations of varying fidelities: for instance, empirically-calibrated semi-analytic emulators \citep[e.g.,][]{takahashi2012revising, mead2016accurate, mead2021hmcode}, fast-executing lognormal dark matter simulations \citep[e.g.,][]{tessore2023glass, lin2023simulation, von2025kids} and ML-based emulators \citep[e.g.,][]{heitmann2009coyote,Knabenhans21,bacco,Giri_2021,piras2023fast}. These techniques could be used to build large mock pre-training datasets, allowing for a significant reduction in the computation time required for the production of high-fidelity simulation datasets for transfer learning. Similarly, computation budgets could be reoriented towards fewer high-fidelity simulations with more particles or larger simulation boxes. Either way, by enabling an order of magnitude reduction in high-fidelity simulations, this work demonstrates that multifidelity transfer learning has the potential to transform our approach to  simulation-based inference in cosmology.

\section*{Acknowledgements}

AAS was supported by the STFC UCL Centre for Doctoral Training in Data Intensive Science (grant ST/W00674X/1) and by departmental and industry contributions. AAS was also supported by the A. G. Leventis Foundation educational grant scheme. DP was supported by a Swiss National Science Foundation (SNSF) grant, and by the SNF Sinergia grant CRSII5-193826 ``AstroSignals: A New Window on the Universe, with the New Generation of Large Radio-Astronomy Facilities''. AMGF is grateful to support from the UPFLOW project, funded by the European Research Council under the European Union's Horizon 2020 research and innovation program (grant agreement No 101001601). NJ and BJ acknowledge support by the ERC-selected UKRI Frontier Research Grant EP/Y03015X/1.

%%%%%%%%%%%%%%%%%%%%%%%%%%%%%%%%%%%%%%%%%%%%%%%%%%
\section*{Data Availability}

The \texttt{Python} software used to produce the results of this paper is available at \href{https://github.com/asaoulis/transfer-sbi}{https://github.com/asaoulis/transfer-sbi}. All the simulation data used is publicly available via the CAMELS Multifield Dataset \citep{villaescusa2021camels, villaescusa2022camels}.

\bibliographystyle{mnras}
\bibliography{example} % if your bibtex file is called example.bib

\begin{thebibliography}{}
\makeatletter
\relax
\def\mn@urlcharsother{\let\do\@makeother \do\$\do\&\do\#\do\^\do\_\do\%\do\~}
\def\mn@doi{\begingroup\mn@urlcharsother \@ifnextchar [ {\mn@doi@} {\mn@doi@[]}}
\def\mn@doi@[#1]#2{\def\@tempa{#1}\ifx\@tempa\@empty \href {http://dx.doi.org/#2} {doi:#2}\else \href {http://dx.doi.org/#2} {#1}\fi \endgroup}
\def\mn@eprint#1#2{\mn@eprint@#1:#2::\@nil}
\def\mn@eprint@arXiv#1{\href {http://arxiv.org/abs/#1} {{\tt arXiv:#1}}}
\def\mn@eprint@dblp#1{\href {http://dblp.uni-trier.de/rec/bibtex/#1.xml} {dblp:#1}}
\def\mn@eprint@#1:#2:#3:#4\@nil{\def\@tempa {#1}\def\@tempb {#2}\def\@tempc {#3}\ifx \@tempc \@empty \let \@tempc \@tempb \let \@tempb \@tempa \fi \ifx \@tempb \@empty \def\@tempb {arXiv}\fi \@ifundefined {mn@eprint@\@tempb}{\@tempb:\@tempc}{\expandafter \expandafter \csname mn@eprint@\@tempb\endcsname \expandafter{\@tempc}}}

\bibitem[\protect\citeauthoryear{Abdalla et~al.,}{Abdalla et~al.}{2022}]{abdalla2022cosmology}
Abdalla E.,  et~al., 2022, Journal of High Energy Astrophysics, 34, 49

\bibitem[\protect\citeauthoryear{Akhmetzhanova, Mishra-Sharma  \& Dvorkin}{Akhmetzhanova et~al.}{2024}]{akhmetzhanova2024data}
Akhmetzhanova A.,  Mishra-Sharma S.,   Dvorkin C.,  2024, Monthly Notices of the Royal Astronomical Society, 527, 7459

\bibitem[\protect\citeauthoryear{Akiba, Sano, Yanase, Ohta  \& Koyama}{Akiba et~al.}{2019}]{akiba2019optuna}
Akiba T.,  Sano S.,  Yanase T.,  Ohta T.,   Koyama M.,  2019, in Proceedings of the 25th ACM SIGKDD international conference on knowledge discovery \& data mining. pp 2623--2631

\bibitem[\protect\citeauthoryear{Alsing, Wandelt  \& Feeney}{Alsing et~al.}{2018}]{alsing2018massive}
Alsing J.,  Wandelt B.,   Feeney S.,  2018, Monthly Notices of the Royal Astronomical Society, 477, 2874

\bibitem[\protect\citeauthoryear{Alsing, Charnock, Feeney  \& Wandelt}{Alsing et~al.}{2019}]{alsing2019fast}
Alsing J.,  Charnock T.,  Feeney S.,   Wandelt B.,  2019, Monthly Notices of the Royal Astronomical Society, 488, 4440

\bibitem[\protect\citeauthoryear{Andrianomena \& Hassan}{Andrianomena \& Hassan}{2023}]{andrianomena2023predictive}
Andrianomena S.,  Hassan S.,  2023, Journal of Cosmology and Astroparticle Physics, 2023, 051

\bibitem[\protect\citeauthoryear{Andrianomena \& Hassan}{Andrianomena \& Hassan}{2025}]{andrianomena2025towards}
Andrianomena S.,  Hassan S.,  2025, Astrophysics and Space Science, 370, 14

\bibitem[\protect\citeauthoryear{Aricò, Angulo, Contreras, Ondaro-Mallea, Pellejero-Ibañez  \& Zennaro}{Aricò et~al.}{2021}]{bacco}
Aricò G.,  Angulo R.~E.,  Contreras S.,  Ondaro-Mallea L.,  Pellejero-Ibañez M.,   Zennaro M.,  2021, \mn@doi [Monthly Notices of the Royal Astronomical Society] {10.1093/mnras/stab1911}, 506, 4070

\bibitem[\protect\citeauthoryear{Bairagi, Wandelt  \& Villaescusa-Navarro}{Bairagi et~al.}{2025}]{bairagi2025many}
Bairagi A.,  Wandelt B.,   Villaescusa-Navarro F.,  2025, arXiv preprint arXiv:2503.13755

\bibitem[\protect\citeauthoryear{Bengio}{Bengio}{2012}]{bengio2012deep}
Bengio Y.,  2012, in Proceedings of ICML workshop on unsupervised and transfer learning. pp 17--36

\bibitem[\protect\citeauthoryear{Bernardini et~al.,}{Bernardini et~al.}{2025}]{bernardini2025ember}
Bernardini M.,  et~al., 2025, Monthly Notices of the Royal Astronomical Society, 538, 1201

\bibitem[\protect\citeauthoryear{Boruah, Rozo  \& Fiedorowicz}{Boruah et~al.}{2022}]{boruah2022map}
Boruah S.~S.,  Rozo E.,   Fiedorowicz P.,  2022, Monthly Notices of the Royal Astronomical Society, 516, 4111

\bibitem[\protect\citeauthoryear{Carlini, Liu, Erlingsson, Kos  \& Song}{Carlini et~al.}{2019}]{carlini2019secret}
Carlini N.,  Liu C.,  Erlingsson {\'U}.,  Kos J.,   Song D.,  2019, in 28th USENIX security symposium (USENIX security 19). pp 267--284

\bibitem[\protect\citeauthoryear{Carlini, Ippolito, Jagielski, Lee, Tramer  \& Zhang}{Carlini et~al.}{2023}]{carlini2022quantifying}
Carlini N.,  Ippolito D.,  Jagielski M.,  Lee K.,  Tramer F.,   Zhang C.,  2023, in The Eleventh International Conference on Learning Representations. \url {https://openreview.net/forum?id=TatRHT_1cK}

\bibitem[\protect\citeauthoryear{Chartier, Wandelt, Akrami  \& Villaescusa-Navarro}{Chartier et~al.}{2021}]{chartier2021carpool}
Chartier N.,  Wandelt B.,  Akrami Y.,   Villaescusa-Navarro F.,  2021, Monthly Notices of the Royal Astronomical Society, 503, 1897

\bibitem[\protect\citeauthoryear{Cheng, Ting, M{\'e}nard  \& Bruna}{Cheng et~al.}{2020}]{cheng2020new}
Cheng S.,  Ting Y.-S.,  M{\'e}nard B.,   Bruna J.,  2020, Monthly Notices of the Royal Astronomical Society, 499, 5902

\bibitem[\protect\citeauthoryear{Cheng, Marques, Grand{\'o}n, Thiele, Shirasaki, M{\'e}nard  \& Liu}{Cheng et~al.}{2025}]{cheng2025cosmological}
Cheng S.,  Marques G.~A.,  Grand{\'o}n D.,  Thiele L.,  Shirasaki M.,  M{\'e}nard B.,   Liu J.,  2025, Journal of Cosmology and Astroparticle Physics, 2025, 006

\bibitem[\protect\citeauthoryear{Cranmer, Brehmer  \& Louppe}{Cranmer et~al.}{2020}]{cranmer2020frontier}
Cranmer K.,  Brehmer J.,   Louppe G.,  2020, Proceedings of the National Academy of Sciences, 117, 30055

\bibitem[\protect\citeauthoryear{Dai \& Seljak}{Dai \& Seljak}{2024}]{dai2024multiscale}
Dai B.,  Seljak U.,  2024, Proceedings of the National Academy of Sciences, 121, e2309624121

\bibitem[\protect\citeauthoryear{Deistler, Goncalves  \& Macke}{Deistler et~al.}{2022}]{deistler2022truncated}
Deistler M.,  Goncalves P.~J.,   Macke J.~H.,  2022, in Advances in Neural Information Processing Systems. Curran Associates, Inc., pp 23135--23149, \url {https://proceedings.neurips.cc/paper_files/paper/2022/file/9278abf072b58caf21d48dd670b4c721-Paper-Conference.pdf}

\bibitem[\protect\citeauthoryear{Delaunoy, Hermans, Rozet, Wehenkel  \& Louppe}{Delaunoy et~al.}{2022}]{delaunoy2022towards}
Delaunoy A.,  Hermans J.,  Rozet F.,  Wehenkel A.,   Louppe G.,  2022, in Advances in Neural Information Processing Systems. Curran Associates, Inc., pp 20025--20037, \url {https://proceedings.neurips.cc/paper_files/paper/2022/file/7e6288bfb68182db7d6e328b0aefa89a-Paper-Conference.pdf}

\bibitem[\protect\citeauthoryear{Delaunoy, Bonardeaux, Mishra-Sharma  \& Louppe}{Delaunoy et~al.}{2024}]{delaunoy2024low}
Delaunoy A.,  Bonardeaux M. d. l.~B.,  Mishra-Sharma S.,   Louppe G.,  2024, arXiv preprint arXiv:2408.15136

\bibitem[\protect\citeauthoryear{Devlin, Chang, Lee  \& Toutanova}{Devlin et~al.}{2019}]{devlin2019bert}
Devlin J.,  Chang M.-W.,  Lee K.,   Toutanova K.,  2019, in Proceedings of the 2019 conference of the North American chapter of the association for computational linguistics: human language technologies, volume 1 (long and short papers). pp 4171--4186

\bibitem[\protect\citeauthoryear{Dosovitskiy et~al.,}{Dosovitskiy et~al.}{2021}]{dosovitskiy2020image}
Dosovitskiy A.,  et~al., 2021, in International Conference on Learning Representations. \url {https://openreview.net/forum?id=YicbFdNTTy}

\bibitem[\protect\citeauthoryear{Durkan, Bekasov, Murray  \& Papamakarios}{Durkan et~al.}{2019}]{durkan2019neural}
Durkan C.,  Bekasov A.,  Murray I.,   Papamakarios G.,  2019, in Advances in Neural Information Processing Systems. Curran Associates, Inc., \url {https://proceedings.neurips.cc/paper_files/paper/2019/file/7ac71d433f282034e088473244df8c02-Paper.pdf}

\bibitem[\protect\citeauthoryear{Durkan, Murray  \& Papamakarios}{Durkan et~al.}{2020}]{durkan2020contrastive}
Durkan C.,  Murray I.,   Papamakarios G.,  2020, in III H.~D.,  Singh A.,  eds,  Proceedings of Machine Learning Research Vol. 119, Proceedings of the 37th International Conference on Machine Learning. PMLR, pp 2771--2781, \url {https://proceedings.mlr.press/v119/durkan20a.html}

\bibitem[\protect\citeauthoryear{Elbers et~al.,}{Elbers et~al.}{2025}]{elbers2025flamingo}
Elbers W.,  et~al., 2025, Monthly Notices of the Royal Astronomical Society, 537, 2160

\bibitem[\protect\citeauthoryear{Euclid Collaboration:~Knabenhans et~al.,}{Euclid Collaboration:~Knabenhans et~al.}{2021}]{Knabenhans21}
Euclid Collaboration:~Knabenhans M.,  et~al., 2021, \mn@doi [Monthly Notices of the Royal Astronomical Society] {10.1093/mnras/stab1366}, 505, 2840

\bibitem[\protect\citeauthoryear{{Euclid Collaboration} et~al.,}{{Euclid Collaboration} et~al.}{2025}]{mellier2024euclid}
{Euclid Collaboration} et~al., 2025, \mn@doi [\aap] {10.1051/0004-6361/202450810}, \href {https://ui.adsabs.harvard.edu/abs/2025A&A...697A...1E} {697, A1}

\bibitem[\protect\citeauthoryear{Fluri, Kacprzak, Lucchi, Refregier, Amara, Hofmann  \& Schneider}{Fluri et~al.}{2019}]{fluri2019cosmological}
Fluri J.,  Kacprzak T.,  Lucchi A.,  Refregier A.,  Amara A.,  Hofmann T.,   Schneider A.,  2019, Physical Review D, 100, 063514

\bibitem[\protect\citeauthoryear{Fluri, Kacprzak, Lucchi, Schneider, Refregier  \& Hofmann}{Fluri et~al.}{2022}]{fluri2022full}
Fluri J.,  Kacprzak T.,  Lucchi A.,  Schneider A.,  Refregier A.,   Hofmann T.,  2022, Physical Review D, 105, 083518

\bibitem[\protect\citeauthoryear{Friedman}{Friedman}{2004}]{friedman2004multivariate}
Friedman J.,  2004, Technical report, On multivariate goodness-of-fit and two-sample testing.
SLAC National Accelerator Laboratory (SLAC), Menlo Park, CA (United States)

\bibitem[\protect\citeauthoryear{Ganin \& Lempitsky}{Ganin \& Lempitsky}{2015}]{ganin2015unsupervised}
Ganin Y.,  Lempitsky V.,  2015, in Bach F.,  Blei D.,  eds,  Proceedings of Machine Learning Research Vol. 37, Proceedings of the 32nd International Conference on Machine Learning. PMLR, Lille, France, pp 1180--1189, \url {https://proceedings.mlr.press/v37/ganin15.html}

\bibitem[\protect\citeauthoryear{Gatti et~al.,}{Gatti et~al.}{2024}]{gatti2024dark}
Gatti M.,  et~al., 2024, Physical Review D, 109, 063534

\bibitem[\protect\citeauthoryear{Gehrels et~al.,}{Gehrels et~al.}{2015}]{spergel2015wide}
Gehrels N.,  et~al., 2015, arXiv preprint arXiv:1503.03757

\bibitem[\protect\citeauthoryear{Giri \& Schneider}{Giri \& Schneider}{2021}]{Giri_2021}
Giri S.~K.,  Schneider A.,  2021, \mn@doi [Journal of Cosmology and Astroparticle Physics] {10.1088/1475-7516/2021/12/046}, 2021, 046

\bibitem[\protect\citeauthoryear{Girshick, Donahue, Darrell  \& Malik}{Girshick et~al.}{2014}]{girshick2014rich}
Girshick R.,  Donahue J.,  Darrell T.,   Malik J.,  2014, in Proceedings of the IEEE conference on computer vision and pattern recognition. pp 580--587

\bibitem[\protect\citeauthoryear{Gluck, Oppenheimer, Nagai, Villaescusa-Navarro  \& Angl{\'e}s-Alc{\'a}zar}{Gluck et~al.}{2024}]{gluck2024observationally}
Gluck N.,  Oppenheimer B.~D.,  Nagai D.,  Villaescusa-Navarro F.,   Angl{\'e}s-Alc{\'a}zar D.,  2024, Monthly Notices of the Royal Astronomical Society, 527, 10038

\bibitem[\protect\citeauthoryear{Gondhalekar \& Moriwaki}{Gondhalekar \& Moriwaki}{2024}]{gondhalekar2024convolutional}
Gondhalekar Y.,  Moriwaki K.,  2024, arXiv preprint arXiv:2411.14392

\bibitem[\protect\citeauthoryear{Gotmare, Keskar, Xiong  \& Socher}{Gotmare et~al.}{2019}]{gotmare2018a}
Gotmare A.,  Keskar N.~S.,  Xiong C.,   Socher R.,  2019, in International Conference on Learning Representations. \url {https://openreview.net/forum?id=r14EOsCqKX}

\bibitem[\protect\citeauthoryear{Goyal et~al.,}{Goyal et~al.}{2017}]{goyal2017accurate}
Goyal P.,  et~al., 2017, arXiv preprint arXiv:1706.02677

\bibitem[\protect\citeauthoryear{Greenberg, Nonnenmacher  \& Macke}{Greenberg et~al.}{2019}]{greenberg2019automatic}
Greenberg D.,  Nonnenmacher M.,   Macke J.,  2019, in Chaudhuri K.,  Salakhutdinov R.,  eds,  Proceedings of Machine Learning Research Vol. 97, Proceedings of the 36th International Conference on Machine Learning. PMLR, pp 2404--2414, \url {https://proceedings.mlr.press/v97/greenberg19a.html}

\bibitem[\protect\citeauthoryear{Gupta, Matilla, Hsu  \& Haiman}{Gupta et~al.}{2018}]{gupta2018non}
Gupta A.,  Matilla J. M.~Z.,  Hsu D.,   Haiman Z.,  2018, Physical Review D, 97, 103515

\bibitem[\protect\citeauthoryear{Hahn et~al.,}{Hahn et~al.}{2024}]{hahn2024cosmological}
Hahn C.,  et~al., 2024, Physical Review D, 109, 083534

\bibitem[\protect\citeauthoryear{Halder, Friedrich, Seitz  \& Varga}{Halder et~al.}{2021}]{halder2021integrated}
Halder A.,  Friedrich O.,  Seitz S.,   Varga T.~N.,  2021, Monthly Notices of the Royal Astronomical Society, 506, 2780

\bibitem[\protect\citeauthoryear{Harnois-D{\'e}raps, Martinet, Castro, Dolag, Giblin, Heymans, Hildebrandt  \& Xia}{Harnois-D{\'e}raps et~al.}{2021}]{harnois2021cosmic}
Harnois-D{\'e}raps J.,  Martinet N.,  Castro T.,  Dolag K.,  Giblin B.,  Heymans C.,  Hildebrandt H.,   Xia Q.,  2021, Monthly Notices of the Royal Astronomical Society, 506, 1623

\bibitem[\protect\citeauthoryear{Harnois-D{\'e}raps et~al.,}{Harnois-D{\'e}raps et~al.}{2024}]{harnois2024kids}
Harnois-D{\'e}raps J.,  et~al., 2024, Monthly Notices of the Royal Astronomical Society, 534, 3305

\bibitem[\protect\citeauthoryear{Hassan, Andrianomena  \& Doughty}{Hassan et~al.}{2020}]{hassan2020constraining}
Hassan S.,  Andrianomena S.,   Doughty C.,  2020, Monthly Notices of the Royal Astronomical Society, 494, 5761

\bibitem[\protect\citeauthoryear{He, Zhang, Ren  \& Sun}{He et~al.}{2016}]{he2016deep}
He K.,  Zhang X.,  Ren S.,   Sun J.,  2016, in Proceedings of the IEEE conference on computer vision and pattern recognition. pp 770--778

\bibitem[\protect\citeauthoryear{He, Chen, Xie, Li, Doll{\'a}r  \& Girshick}{He et~al.}{2022}]{he2022masked}
He K.,  Chen X.,  Xie S.,  Li Y.,  Doll{\'a}r P.,   Girshick R.,  2022, in Proceedings of the IEEE/CVF conference on computer vision and pattern recognition. pp 16000--16009

\bibitem[\protect\citeauthoryear{Heitmann, Higdon, White, Habib, Williams, Lawrence  \& Wagner}{Heitmann et~al.}{2009}]{heitmann2009coyote}
Heitmann K.,  Higdon D.,  White M.,  Habib S.,  Williams B.~J.,  Lawrence E.,   Wagner C.,  2009, The Astrophysical Journal, 705, 156

\bibitem[\protect\citeauthoryear{Hermans, Begy  \& Louppe}{Hermans et~al.}{2020}]{hermans2020likelihood}
Hermans J.,  Begy V.,   Louppe G.,  2020, in III H.~D.,  Singh A.,  eds,  Proceedings of Machine Learning Research Vol. 119, Proceedings of the 37th International Conference on Machine Learning. PMLR, pp 4239--4248, \url {https://proceedings.mlr.press/v119/hermans20a.html}

\bibitem[\protect\citeauthoryear{Hermans, Delaunoy, Rozet, Wehenkel, Begy  \& Louppe}{Hermans et~al.}{2022}]{hermans2022crisis}
Hermans J.,  Delaunoy A.,  Rozet F.,  Wehenkel A.,  Begy V.,   Louppe G.,  2022, Transactions on Machine Learning Research, p. https://openreview.net/forum?id=LHAbHkt6Aq

\bibitem[\protect\citeauthoryear{Hikida, Bharti, Jeffrey  \& Briol}{Hikida et~al.}{2025}]{hikida2025multilevel}
Hikida Y.,  Bharti A.,  Jeffrey N.,   Briol F.-X.,  2025, arXiv preprint arXiv:2506.06087

\bibitem[\protect\citeauthoryear{Hoffmann, Bar-Sinai, Lee, Andrejevic, Mishra, Rubinstein  \& Rycroft}{Hoffmann et~al.}{2019}]{hoffmann2019machine}
Hoffmann J.,  Bar-Sinai Y.,  Lee L.~M.,  Andrejevic J.,  Mishra S.,  Rubinstein S.~M.,   Rycroft C.~H.,  2019, Science advances, 5, eaau6792

\bibitem[\protect\citeauthoryear{Ioffe \& Szegedy}{Ioffe \& Szegedy}{2015}]{ioffe2015batch}
Ioffe S.,  Szegedy C.,  2015, in Bach F.,  Blei D.,  eds,  Proceedings of Machine Learning Research Vol. 37, Proceedings of the 32nd International Conference on Machine Learning. PMLR, Lille, France, pp 448--456, \url {https://proceedings.mlr.press/v37/ioffe15.html}

\bibitem[\protect\citeauthoryear{Ivezi{\'c} et~al.,}{Ivezi{\'c} et~al.}{2019}]{ivezic2019lsst}
Ivezi{\'c} {\v{Z}}.,  et~al., 2019, The Astrophysical Journal, 873, 111

\bibitem[\protect\citeauthoryear{Jarvis, Bernstein  \& Jain}{Jarvis et~al.}{2004}]{jarvis2004skewness}
Jarvis M.,  Bernstein G.,   Jain B.,  2004, Monthly Notices of the Royal Astronomical Society, 352, 338

\bibitem[\protect\citeauthoryear{Jasche \& Lavaux}{Jasche \& Lavaux}{2019}]{jasche2019physical}
Jasche J.,  Lavaux G.,  2019, Astronomy \& Astrophysics, 625, A64

\bibitem[\protect\citeauthoryear{Jasche \& Wandelt}{Jasche \& Wandelt}{2013}]{jasche2013bayesian}
Jasche J.,  Wandelt B.~D.,  2013, Monthly Notices of the Royal Astronomical Society, 432, 894

\bibitem[\protect\citeauthoryear{Jasche, Leclercq  \& Wandelt}{Jasche et~al.}{2015}]{jasche2015past}
Jasche J.,  Leclercq F.,   Wandelt B.~D.,  2015, Journal of Cosmology and Astroparticle Physics, 2015, 036

\bibitem[\protect\citeauthoryear{Jeffrey, Alsing  \& Lanusse}{Jeffrey et~al.}{2021}]{jeffrey2021likelihood}
Jeffrey N.,  Alsing J.,   Lanusse F.,  2021, Monthly Notices of the Royal Astronomical Society, 501, 954

\bibitem[\protect\citeauthoryear{Jeffrey et~al.,}{Jeffrey et~al.}{2025}]{jeffrey2025dark}
Jeffrey N.,  et~al., 2025, Monthly Notices of the Royal Astronomical Society, 536, 1303

\bibitem[\protect\citeauthoryear{Jia}{Jia}{2024a}]{jia2024cosmological}
Jia H.,  2024a, arXiv preprint arXiv:2411.14748

\bibitem[\protect\citeauthoryear{Jia}{Jia}{2024b}]{jia2024simulation}
Jia H.,  2024b, in Salakhutdinov R.,  Kolter Z.,  Heller K.,  Weller A.,  Oliver N.,  Scarlett J.,   Berkenkamp F.,  eds,  Proceedings of Machine Learning Research Vol. 235, Proceedings of the 41st International Conference on Machine Learning. PMLR, pp 21731--21752, \url {https://proceedings.mlr.press/v235/jia24a.html}

\bibitem[\protect\citeauthoryear{Jo, Genel, Sengupta, Wandelt, Somerville  \& Villaescusa-Navarro}{Jo et~al.}{2025}]{jo2025towards}
Jo Y.,  Genel S.,  Sengupta A.,  Wandelt B.,  Somerville R.,   Villaescusa-Navarro F.,  2025, arXiv preprint arXiv:2502.13239

\bibitem[\protect\citeauthoryear{Kalra \& Barkeshli}{Kalra \& Barkeshli}{2024}]{kalra2024warmup}
Kalra D.~S.,  Barkeshli M.,  2024, in Advances in Neural Information Processing Systems. Curran Associates, Inc., pp 111760--111801, \url {https://proceedings.neurips.cc/paper_files/paper/2024/file/ca98452d4e9ecbc18c40da2aa0da8b98-Paper-Conference.pdf}

\bibitem[\protect\citeauthoryear{Kirillov et~al.,}{Kirillov et~al.}{2023}]{kirillov2023segment}
Kirillov A.,  et~al., 2023, in Proceedings of the IEEE/CVF international conference on computer vision. pp 4015--4026

\bibitem[\protect\citeauthoryear{Kornblith, Shlens  \& Le}{Kornblith et~al.}{2019}]{kornblith2019better}
Kornblith S.,  Shlens J.,   Le Q.~V.,  2019, in Proceedings of the IEEE/CVF conference on computer vision and pattern recognition. pp 2661--2671

\bibitem[\protect\citeauthoryear{Krogh \& Hertz}{Krogh \& Hertz}{1991}]{krogh1991simple}
Krogh A.,  Hertz J.,  1991, in Advances in Neural Information Processing Systems. Morgan-Kaufmann, \url {https://proceedings.neurips.cc/paper_files/paper/1991/file/8eefcfdf5990e441f0fb6f3fad709e21-Paper.pdf}

\bibitem[\protect\citeauthoryear{Krouglova, Johnson, Confavreux, Deistler  \& Gon{\c{c}}alves}{Krouglova et~al.}{2025}]{krouglova2025multifidelity}
Krouglova A.~N.,  Johnson H.~R.,  Confavreux B.,  Deistler M.,   Gon{\c{c}}alves P.~J.,  2025, arXiv preprint arXiv:2502.08416

\bibitem[\protect\citeauthoryear{Lanzieri, Zeghal, Makinen, Boucaud, Starck  \& Lanusse}{Lanzieri et~al.}{2024}]{lanzieri2024optimal}
Lanzieri D.,  Zeghal J.,  Makinen T.~L.,  Boucaud A.,  Starck J.-L.,   Lanusse F.,  2024, arXiv preprint arXiv:2407.10877

\bibitem[\protect\citeauthoryear{{Lastufka} et~al.,}{{Lastufka} et~al.}{2024}]{Lastufka24}
{Lastufka} E.,  et~al., 2024, \mn@doi [arXiv e-prints] {10.48550/arXiv.2409.11175}, \href {https://ui.adsabs.harvard.edu/abs/2024arXiv240911175L} {p. arXiv:2409.11175}

\bibitem[\protect\citeauthoryear{Leclercq \& Heavens}{Leclercq \& Heavens}{2021}]{leclercq2021accuracy}
Leclercq F.,  Heavens A.,  2021, Monthly Notices of the Royal Astronomical Society: Letters, 506, L85

\bibitem[\protect\citeauthoryear{Lee et~al.,}{Lee et~al.}{2024}]{lee2024zooming}
Lee M.~E.,  et~al., 2024, The Astrophysical Journal, 968, 11

\bibitem[\protect\citeauthoryear{Lemos, Cranmer, Abidi, Hahn, Eickenberg, Massara, Yallup  \& Ho}{Lemos et~al.}{2023a}]{lemos2023robust}
Lemos P.,  Cranmer M.,  Abidi M.,  Hahn C.,  Eickenberg M.,  Massara E.,  Yallup D.,   Ho S.,  2023a, Machine Learning: Science and Technology, 4, 01LT01

\bibitem[\protect\citeauthoryear{Lemos, Coogan, Hezaveh  \& Perreault-Levasseur}{Lemos et~al.}{2023b}]{lemos2023sampling}
Lemos P.,  Coogan A.,  Hezaveh Y.,   Perreault-Levasseur L.,  2023b, in Proceedings of the 40th International Conference on Machine Learning. PMLR, pp 19256--19273, \url {https://proceedings.mlr.press/v202/lemos23a.html}

\bibitem[\protect\citeauthoryear{Lemos et~al.,}{Lemos et~al.}{2024}]{lemos2024field}
Lemos P.,  et~al., 2024, Physical Review D, 109, 083536

\bibitem[\protect\citeauthoryear{Lian, Lindblad, Micke  \& Sladoje}{Lian et~al.}{2025}]{lian2025isolated}
Lian W.,  Lindblad J.,  Micke P.,   Sladoje N.,  2025, arXiv preprint arXiv:2503.09826

\bibitem[\protect\citeauthoryear{Lin, von Wietersheim-Kramsta, Joachimi  \& Feeney}{Lin et~al.}{2023}]{lin2023simulation}
Lin K.,  von Wietersheim-Kramsta M.,  Joachimi B.,   Feeney S.,  2023, Monthly Notices of the Royal Astronomical Society, 524, 6167

\bibitem[\protect\citeauthoryear{Liu, Mao, Wu, Feichtenhofer, Darrell  \& Xie}{Liu et~al.}{2022}]{liu2022convnet}
Liu Z.,  Mao H.,  Wu C.-Y.,  Feichtenhofer C.,  Darrell T.,   Xie S.,  2022, in Proceedings of the IEEE/CVF conference on computer vision and pattern recognition. pp 11976--11986

\bibitem[\protect\citeauthoryear{Lopez-Paz \& Oquab}{Lopez-Paz \& Oquab}{2017}]{lopez2017revisiting}
Lopez-Paz D.,  Oquab M.,  2017, in International Conference on Learning Representations. \url {https://openreview.net/forum?id=SJkXfE5xx}

\bibitem[\protect\citeauthoryear{Loshchilov \& Hutter}{Loshchilov \& Hutter}{2017}]{loshchilov2017decoupled}
Loshchilov I.,  Hutter F.,  2017, arXiv preprint arXiv:1711.05101

\bibitem[\protect\citeauthoryear{Lu, Haiman  \& Li}{Lu et~al.}{2023}]{lu2023cosmological}
Lu T.,  Haiman Z.,   Li X.,  2023, Monthly Notices of the Royal Astronomical Society, 521, 2050

\bibitem[\protect\citeauthoryear{Lueckmann, Bassetto, Karaletsos  \& Macke}{Lueckmann et~al.}{2019}]{lueckmann2019likelihood}
Lueckmann J.-M.,  Bassetto G.,  Karaletsos T.,   Macke J.~H.,  2019, in Symposium on Advances in Approximate Bayesian Inference. pp 32--53

\bibitem[\protect\citeauthoryear{Lueckmann, Boelts, Greenberg, Goncalves  \& Macke}{Lueckmann et~al.}{2021}]{lueckmann2021benchmarking}
Lueckmann J.-M.,  Boelts J.,  Greenberg D.,  Goncalves P.,   Macke J.,  2021, in International conference on artificial intelligence and statistics. pp 343--351

\bibitem[\protect\citeauthoryear{Maas, Hannun, Ng  et~al.}{Maas et~al.}{2013}]{maas2013rectifier}
Maas A.~L.,  Hannun A.~Y.,  Ng A.~Y.,   et~al., 2013, in Proceedings of the 30th International Conference on Machine Learning. \url {https://ai.stanford.edu/~amaas/papers/relu_hybrid_icml2013_final.pdf}

\bibitem[\protect\citeauthoryear{Makinen, Charnock, Alsing  \& Wandelt}{Makinen et~al.}{2021}]{makinen2021lossless}
Makinen T.~L.,  Charnock T.,  Alsing J.,   Wandelt B.~D.,  2021, Journal of Cosmology and Astroparticle Physics, 2021, 049

\bibitem[\protect\citeauthoryear{Mao, Liu, Liu, Li, Shen  \& Wang}{Mao et~al.}{2023}]{mao2023intriguing}
Mao X.,  Liu Y.,  Liu F.,  Li Q.,  Shen W.,   Wang Y.,  2023, in Proceedings of the AAAI Conference on Artificial Intelligence. pp 1905--1913

\bibitem[\protect\citeauthoryear{Martinet, Harnois-D{\'e}raps, Jullo  \& Schneider}{Martinet et~al.}{2021}]{martinet2021probing}
Martinet N.,  Harnois-D{\'e}raps J.,  Jullo E.,   Schneider P.,  2021, Astronomy \& Astrophysics, 646, A62

\bibitem[\protect\citeauthoryear{Matilla, Sharma, Hsu  \& Haiman}{Matilla et~al.}{2020}]{matilla2020interpreting}
Matilla J. M.~Z.,  Sharma M.,  Hsu D.,   Haiman Z.,  2020, Physical Review D, 102, 123506

\bibitem[\protect\citeauthoryear{McCarthy, Bird, Schaye, Harnois-Deraps, Font  \& Van~Waerbeke}{McCarthy et~al.}{2018}]{mccarthy2018bahamas}
McCarthy I.~G.,  Bird S.,  Schaye J.,  Harnois-Deraps J.,  Font A.~S.,   Van~Waerbeke L.,  2018, Monthly Notices of the Royal Astronomical Society, 476, 2999

\bibitem[\protect\citeauthoryear{Mead, Heymans, Lombriser, Peacock, Steele  \& Winther}{Mead et~al.}{2016}]{mead2016accurate}
Mead A.,  Heymans C.,  Lombriser L.,  Peacock J.,  Steele O.,   Winther H.,  2016, Monthly Notices of the Royal Astronomical Society, 459, 1468

\bibitem[\protect\citeauthoryear{Mead, Brieden, Tr{\"o}ster  \& Heymans}{Mead et~al.}{2021}]{mead2021hmcode}
Mead A.,  Brieden S.,  Tr{\"o}ster T.,   Heymans C.,  2021, Monthly Notices of the Royal Astronomical Society, 502, 1401

\bibitem[\protect\citeauthoryear{Mishra, Panda, Phoo, Chen, Karlinsky, Saenko, Saligrama  \& Feris}{Mishra et~al.}{2022}]{mishra2022task2sim}
Mishra S.,  Panda R.,  Phoo C.~P.,  Chen C.-F.~R.,  Karlinsky L.,  Saenko K.,  Saligrama V.,   Feris R.~S.,  2022, in Proceedings of the IEEE/CVF conference on computer vision and pattern recognition. pp 9194--9204

\bibitem[\protect\citeauthoryear{Ni et~al.,}{Ni et~al.}{2023}]{ni2023camels}
Ni Y.,  et~al., 2023, The Astrophysical Journal, 959, 136

\bibitem[\protect\citeauthoryear{Novaes et~al.,}{Novaes et~al.}{2025}]{novaes2025cosmology}
Novaes C.~P.,  et~al., 2025, Physical Review D, 111, 083510

\bibitem[\protect\citeauthoryear{Oquab et~al.,}{Oquab et~al.}{2024}]{oquab2024dinov2}
Oquab M.,  et~al., 2024, Transactions on Machine Learning Research Journal, pp 1--31

\bibitem[\protect\citeauthoryear{Papamakarios \& Murray}{Papamakarios \& Murray}{2016}]{papamakarios2016fast}
Papamakarios G.,  Murray I.,  2016, in Advances in Neural Information Processing Systems. Curran Associates, Inc., \url {https://proceedings.neurips.cc/paper_files/paper/2016/file/6aca97005c68f1206823815f66102863-Paper.pdf}

\bibitem[\protect\citeauthoryear{Papamakarios, Pavlakou  \& Murray}{Papamakarios et~al.}{2017}]{papamakarios2017masked}
Papamakarios G.,  Pavlakou T.,   Murray I.,  2017, in Advances in Neural Information Processing Systems. Curran Associates, Inc., \url {https://proceedings.neurips.cc/paper_files/paper/2017/file/6c1da886822c67822bcf3679d04369fa-Paper.pdf}

\bibitem[\protect\citeauthoryear{Papamakarios, Sterratt  \& Murray}{Papamakarios et~al.}{2019}]{papamakarios2019sequential}
Papamakarios G.,  Sterratt D.,   Murray I.,  2019, in Chaudhuri K.,  Sugiyama M.,  eds,  Proceedings of Machine Learning Research Vol. 89, Proceedings of the Twenty-Second International Conference on Artificial Intelligence and Statistics. PMLR, pp 837--848, \url {https://proceedings.mlr.press/v89/papamakarios19a.html}

\bibitem[\protect\citeauthoryear{Park, Gatti  \& Jain}{Park et~al.}{2025}]{park2025dimensionality}
Park M.,  Gatti M.,   Jain B.,  2025, Physical Review D, 111, 063523

\bibitem[\protect\citeauthoryear{Pillepich et~al.,}{Pillepich et~al.}{2018}]{pillepich2018simulating}
Pillepich A.,  et~al., 2018, Monthly Notices of the Royal Astronomical Society, 473, 4077

\bibitem[\protect\citeauthoryear{Piras, Joachimi  \& Villaescusa-Navarro}{Piras et~al.}{2023}]{piras2023fast}
Piras D.,  Joachimi B.,   Villaescusa-Navarro F.,  2023, Monthly Notices of the Royal Astronomical Society, 520, 668

\bibitem[\protect\citeauthoryear{Prelogovi{\'c}, Mesinger, Murray, Fiameni  \& Gillet}{Prelogovi{\'c} et~al.}{2022}]{prelogovic2022machine}
Prelogovi{\'c} D.,  Mesinger A.,  Murray S.,  Fiameni G.,   Gillet N.,  2022, Monthly Notices of the Royal Astronomical Society, 509, 3852

\bibitem[\protect\citeauthoryear{Prescott \& Baker}{Prescott \& Baker}{2020}]{prescott2020multifidelity}
Prescott T.~P.,  Baker R.~E.,  2020, SIAM/ASA Journal on Uncertainty Quantification, 8, 114

\bibitem[\protect\citeauthoryear{Prescott \& Baker}{Prescott \& Baker}{2021}]{prescott2021multifidelity}
Prescott T.~P.,  Baker R.~E.,  2021, SIAM/ASA Journal on Uncertainty Quantification, 9, 788

\bibitem[\protect\citeauthoryear{Prescott, Warne  \& Baker}{Prescott et~al.}{2024}]{prescott2024efficient}
Prescott T.~P.,  Warne D.~J.,   Baker R.~E.,  2024, Journal of Computational Physics, 496, 112577

\bibitem[\protect\citeauthoryear{Radford et~al.,}{Radford et~al.}{2021}]{radford2021learning}
Radford A.,  et~al., 2021, in Proceedings of the 38th International Conference on Machine Learning. PMLR, pp 8748--8763, \url {https://proceedings.mlr.press/v139/radford21a.html}

\bibitem[\protect\citeauthoryear{R{\'e}galdo-Saint~Blancard et~al.,}{R{\'e}galdo-Saint~Blancard et~al.}{2024}]{regaldo2024galaxy}
R{\'e}galdo-Saint~Blancard B.,  et~al., 2024, Physical Review D, 109, 083535

\bibitem[\protect\citeauthoryear{Ribli, Pataki, Zorrilla~Matilla, Hsu, Haiman  \& Csabai}{Ribli et~al.}{2019}]{ribli2019weak}
Ribli D.,  Pataki B.~{\'A}.,  Zorrilla~Matilla J.~M.,  Hsu D.,  Haiman Z.,   Csabai I.,  2019, Monthly Notices of the Royal Astronomical Society, 490, 1843

\bibitem[\protect\citeauthoryear{Roncoli, {\'C}iprijanovi{\'c}, Voetberg, Villaescusa-Navarro  \& Nord}{Roncoli et~al.}{2023}]{roncoli2023domain}
Roncoli A.,  {\'C}iprijanovi{\'c} A.,  Voetberg M.,  Villaescusa-Navarro F.,   Nord B.,  2023, arXiv preprint arXiv:2311.01588

\bibitem[\protect\citeauthoryear{Santurkar, Tsipras, Ilyas  \& Madry}{Santurkar et~al.}{2018}]{santurkar2018does}
Santurkar S.,  Tsipras D.,  Ilyas A.,   Madry A.,  2018, in Advances in Neural Information Processing Systems. Curran Associates, Inc., \url {https://proceedings.neurips.cc/paper_files/paper/2018/file/905056c1ac1dad141560467e0a99e1cf-Paper.pdf}

\bibitem[\protect\citeauthoryear{Schaye et~al.,}{Schaye et~al.}{2023}]{schaye2023flamingo}
Schaye J.,  et~al., 2023, Monthly Notices of the Royal Astronomical Society, 526, 4978

\bibitem[\protect\citeauthoryear{Schneider \& Lombardi}{Schneider \& Lombardi}{2003}]{schneider2003three}
Schneider P.,  Lombardi M.,  2003, Astronomy \& Astrophysics, 397, 809

\bibitem[\protect\citeauthoryear{Schneider, Van~Waerbeke, Jain  \& Kruse}{Schneider et~al.}{1998}]{schneider1998new}
Schneider P.,  Van~Waerbeke L.,  Jain B.,   Kruse G.,  1998, Monthly Notices of the Royal Astronomical Society, 296, 873

\bibitem[\protect\citeauthoryear{Schneider, Teyssier, Stadel, Chisari, Le~Brun, Amara  \& Refregier}{Schneider et~al.}{2019}]{schneider2019quantifying}
Schneider A.,  Teyssier R.,  Stadel J.,  Chisari N.~E.,  Le~Brun A.~M.,  Amara A.,   Refregier A.,  2019, Journal of Cosmology and Astroparticle Physics, 2019, 020

\bibitem[\protect\citeauthoryear{Schneider, Stoira, Refregier, Weiss, Knabenhans, Stadel  \& Teyssier}{Schneider et~al.}{2020}]{schneider2020baryonic}
Schneider A.,  Stoira N.,  Refregier A.,  Weiss A.~J.,  Knabenhans M.,  Stadel J.,   Teyssier R.,  2020, Journal of Cosmology and Astroparticle Physics, 2020, 019

\bibitem[\protect\citeauthoryear{Secco et~al.,}{Secco et~al.}{2022}]{secco2022dark}
Secco L.~F.,  et~al., 2022, Physical Review D, 105, 103537

\bibitem[\protect\citeauthoryear{Semboloni, Schrabback, van Waerbeke, Vafaei, Hartlap  \& Hilbert}{Semboloni et~al.}{2011}]{semboloni2011weak}
Semboloni E.,  Schrabback T.,  van Waerbeke L.,  Vafaei S.,  Hartlap J.,   Hilbert S.,  2011, Monthly Notices of the Royal Astronomical Society, 410, 143

\bibitem[\protect\citeauthoryear{Sharma, Dai  \& Seljak}{Sharma et~al.}{2024}]{sharma2024comparative}
Sharma D.,  Dai B.,   Seljak U.,  2024, Journal of Cosmology and Astroparticle Physics, 2024, 010

\bibitem[\protect\citeauthoryear{Smith}{Smith}{2017}]{smith2017cyclical}
Smith L.~N.,  2017, in 2017 IEEE winter conference on applications of computer vision (WACV). pp 464--472

\bibitem[\protect\citeauthoryear{Springel}{Springel}{2005}]{springel2005cosmological}
Springel V.,  2005, Monthly notices of the royal astronomical society, 364, 1105

\bibitem[\protect\citeauthoryear{Springel}{Springel}{2010}]{springel2010pur}
Springel V.,  2010, Monthly Notices of the Royal Astronomical Society, 401, 791

\bibitem[\protect\citeauthoryear{Tahir, Ganguli  \& Rotskoff}{Tahir et~al.}{2024}]{tahir2024features}
Tahir J.,  Ganguli S.,   Rotskoff G.~M.,  2024, arXiv preprint arXiv:2410.08194

\bibitem[\protect\citeauthoryear{Takada \& Jain}{Takada \& Jain}{2003}]{takada2003three}
Takada M.,  Jain B.,  2003, Monthly Notices of the Royal Astronomical Society, 340, 580

\bibitem[\protect\citeauthoryear{Takahashi, Sato, Nishimichi, Taruya  \& Oguri}{Takahashi et~al.}{2012}]{takahashi2012revising}
Takahashi R.,  Sato M.,  Nishimichi T.,  Taruya A.,   Oguri M.,  2012, The Astrophysical Journal, 761, 152

\bibitem[\protect\citeauthoryear{Tejero-Cantero, Boelts, Deistler, Lueckmann, Durkan, Gonçalves, Greenberg  \& Macke}{Tejero-Cantero et~al.}{2022}]{tejerosbi2022code}
Tejero-Cantero A.,  Boelts J.,  Deistler M.,  Lueckmann J.-M.,  Durkan C.,  Gonçalves P.,  Greenberg D.,   Macke J.,  2022, {sbi: Simulation-based inference toolkit}, \url {https://github.com/mackelab/sbi}

\bibitem[\protect\citeauthoryear{Tessore, Loureiro, Joachimi, von Wietersheim-Kramsta  \& Jeffrey}{Tessore et~al.}{2023}]{tessore2023glass}
Tessore N.,  Loureiro A.,  Joachimi B.,  von Wietersheim-Kramsta M.,   Jeffrey N.,  2023, The Open Journal of Astrophysics, 6

\bibitem[\protect\citeauthoryear{Thiele, Massara, Pisani, Hahn, Spergel, Ho  \& Wandelt}{Thiele et~al.}{2024}]{thiele2024neutrino}
Thiele L.,  Massara E.,  Pisani A.,  Hahn C.,  Spergel D.~N.,  Ho S.,   Wandelt B.,  2024, The Astrophysical Journal, 969, 89

\bibitem[\protect\citeauthoryear{Thiele, Bayer  \& Takeishi}{Thiele et~al.}{2025}]{thiele2025simulation}
Thiele L.,  Bayer A.~E.,   Takeishi N.,  2025, Simulation-Efficient Cosmological Inference with Multi-Fidelity SBI (\mn@eprint {arXiv} {2507.00514}), \url {https://arxiv.org/abs/2507.00514}

\bibitem[\protect\citeauthoryear{Tucci \& Schmidt}{Tucci \& Schmidt}{2024}]{tucci2024eftoflss}
Tucci B.,  Schmidt F.,  2024, Journal of Cosmology and Astroparticle Physics, 2024, 063

\bibitem[\protect\citeauthoryear{Vaswani, Shazeer, Parmar, Uszkoreit, Jones, Gomez, Kaiser  \& Polosukhin}{Vaswani et~al.}{2017}]{vaswani2017attention}
Vaswani A.,  Shazeer N.,  Parmar N.,  Uszkoreit J.,  Jones L.,  Gomez A.~N.,  Kaiser L.~u.,   Polosukhin I.,  2017, in Advances in Neural Information Processing Systems. Curran Associates, Inc., \url {https://proceedings.neurips.cc/paper_files/paper/2017/file/3f5ee243547dee91fbd053c1c4a845aa-Paper.pdf}

\bibitem[\protect\citeauthoryear{Villaescusa-Navarro et~al.,}{Villaescusa-Navarro et~al.}{2021}]{villaescusa2021camels}
Villaescusa-Navarro F.,  et~al., 2021, The Astrophysical Journal, 915, 71

\bibitem[\protect\citeauthoryear{Villaescusa-Navarro et~al.,}{Villaescusa-Navarro et~al.}{2022}]{villaescusa2022camels}
Villaescusa-Navarro F.,  et~al., 2022, The Astrophysical Journal Supplement Series, 259, 61

\bibitem[\protect\citeauthoryear{Warne, Prescott, Baker  \& Simpson}{Warne et~al.}{2022}]{warne2022multifidelity}
Warne D.~J.,  Prescott T.~P.,  Baker R.~E.,   Simpson M.~J.,  2022, Journal of Computational Physics, 469, 111543

\bibitem[\protect\citeauthoryear{Wehenkel, Gamella, Sener, Behrmann, Sapiro, Cuturi  \& Jacobsen}{Wehenkel et~al.}{2024}]{wehenkel2024addressing}
Wehenkel A.,  Gamella J.~L.,  Sener O.,  Behrmann J.,  Sapiro G.,  Cuturi M.,   Jacobsen J.-H.,  2024, arXiv preprint arXiv:2405.08719

\bibitem[\protect\citeauthoryear{Weinberger et~al.,}{Weinberger et~al.}{2016}]{weinberger2016simulating}
Weinberger R.,  et~al., 2016, Monthly Notices of the Royal Astronomical Society, 465, 3291

\bibitem[\protect\citeauthoryear{Xavier, Abdalla  \& Joachimi}{Xavier et~al.}{2016}]{xavier2016improving}
Xavier H.~S.,  Abdalla F.~B.,   Joachimi B.,  2016, Monthly Notices of the Royal Astronomical Society, 459, 3693

\bibitem[\protect\citeauthoryear{Yang \& Soatto}{Yang \& Soatto}{2020}]{yang2020fda}
Yang Y.,  Soatto S.,  2020, in Proceedings of the IEEE/CVF conference on computer vision and pattern recognition. pp 4085--4095

\bibitem[\protect\citeauthoryear{Yeom, Giacomelli, Fredrikson  \& Jha}{Yeom et~al.}{2018}]{yeom2018privacy}
Yeom S.,  Giacomelli I.,  Fredrikson M.,   Jha S.,  2018, in 2018 IEEE 31st computer security foundations symposium (CSF). pp 268--282

\bibitem[\protect\citeauthoryear{Yosinski, Clune, Bengio  \& Lipson}{Yosinski et~al.}{2014}]{yosinski2014transferable}
Yosinski J.,  Clune J.,  Bengio Y.,   Lipson H.,  2014, in Advances in Neural Information Processing Systems. Curran Associates, Inc., \url {https://proceedings.neurips.cc/paper_files/paper/2014/file/532a2f85b6977104bc93f8580abbb330-Paper.pdf}

\bibitem[\protect\citeauthoryear{Zhai, Kolesnikov, Houlsby  \& Beyer}{Zhai et~al.}{2022}]{zhai2022scaling}
Zhai X.,  Kolesnikov A.,  Houlsby N.,   Beyer L.,  2022, in Proceedings of the IEEE/CVF conference on computer vision and pattern recognition. pp 12104--12113

\bibitem[\protect\citeauthoryear{Zhuang, Qi, Duan, Xi, Zhu, Zhu, Xiong  \& He}{Zhuang et~al.}{2020}]{zhuang2020comprehensive}
Zhuang F.,  Qi Z.,  Duan K.,  Xi D.,  Zhu Y.,  Zhu H.,  Xiong H.,   He Q.,  2020, Proceedings of the IEEE, 109, 43

\bibitem[\protect\citeauthoryear{Z{\"u}rcher et~al.,}{Z{\"u}rcher et~al.}{2022}]{zurcher2022dark}
Z{\"u}rcher D.,  et~al., 2022, Monthly Notices of the Royal Astronomical Society, 511, 2075

\bibitem[\protect\citeauthoryear{von Wietersheim-Kramsta, Lin, Tessore, Joachimi, Loureiro, Reischke  \& Wright}{von Wietersheim-Kramsta et~al.}{2025}]{von2025kids}
von Wietersheim-Kramsta M.,  Lin K.,  Tessore N.,  Joachimi B.,  Loureiro A.,  Reischke R.,   Wright A.~H.,  2025, Astronomy \& Astrophysics, 694, A223

\makeatother
\end{thebibliography}

%%%%%%%%%%%%%%%%%%%%%%%%%%%%%%%%%%%%%%%%%%%%%%%%%%

%%%%%%%%%%%%%%%%% APPENDICES %%%%%%%%%%%%%%%%%%%%%

\appendix

\section{Further posterior comparisons}
\label{app:further_comparison}
\begin{figure*}

	\includegraphics[width=\textwidth]{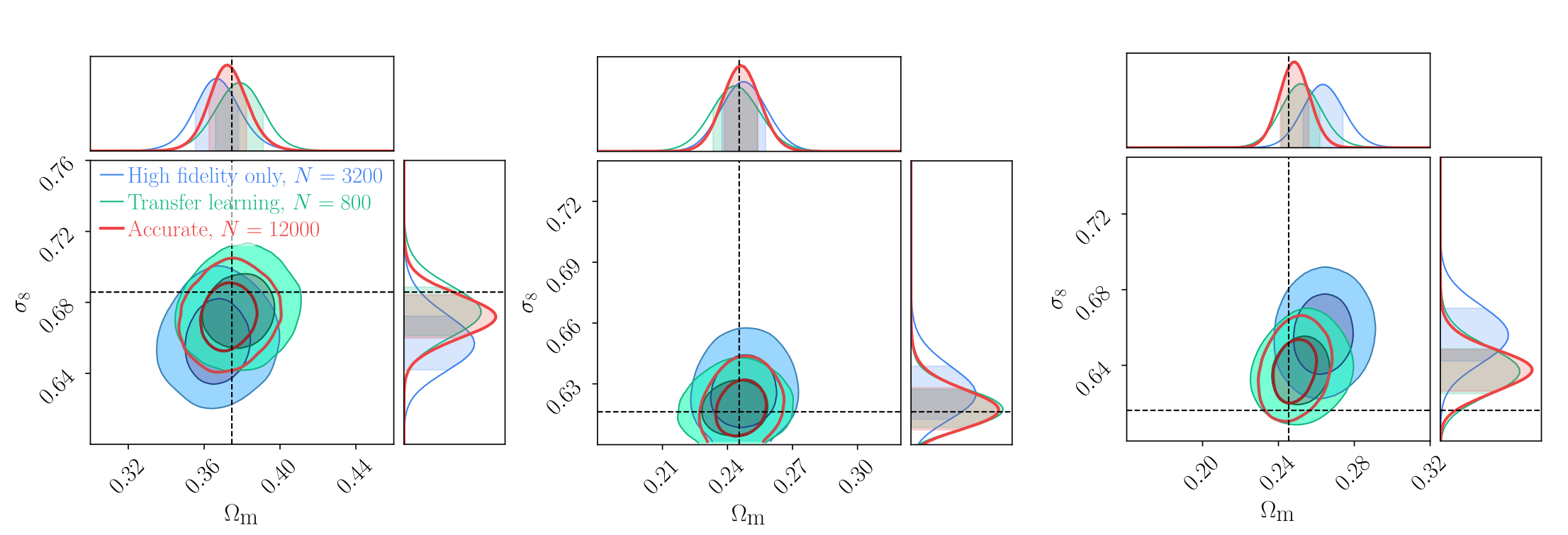}
    \caption{Three representative examples of inference from the LH simulation suite. The true cosmology is shown by the black dashed line. A model trained using transfer learning with $N=800$ high-fidelity IllustrisTNG maps is compared against a high-fidelity-only model trained with $N=3200$ maps. The posteriors are compared with an ``accurate'' posterior model that was trained using the full simulation suite. }
    \label{fig:LH_inference_posteriors_3200}
    
\end{figure*}

We present a further comparison between high-fidelity-only training, with $N=3200$, and the transfer learning approach for the LH suite 2-parameter inference problem in \cref{fig:LH_inference_posteriors_3200}. Despite the factor of $\times4$ increase in the number of high-fidelity maps used during training, the multifidelity approach yields significantly tighter posteriors that better match the ``accurate'' model. In addition, the right-most panel in \cref{fig:LH_inference_posteriors_3200} is suggestive of the overconfidence issue identified in \cref{sec:LH_results}; high-fidelity-only models trained with fewer than $N=6400$ IllustrisTNG maps exhibit a large degree of overconfidence, per \cref{fig:calibration_curves}.

\section{Probing model performance}

\begin{figure*}

	\includegraphics[width=\textwidth]{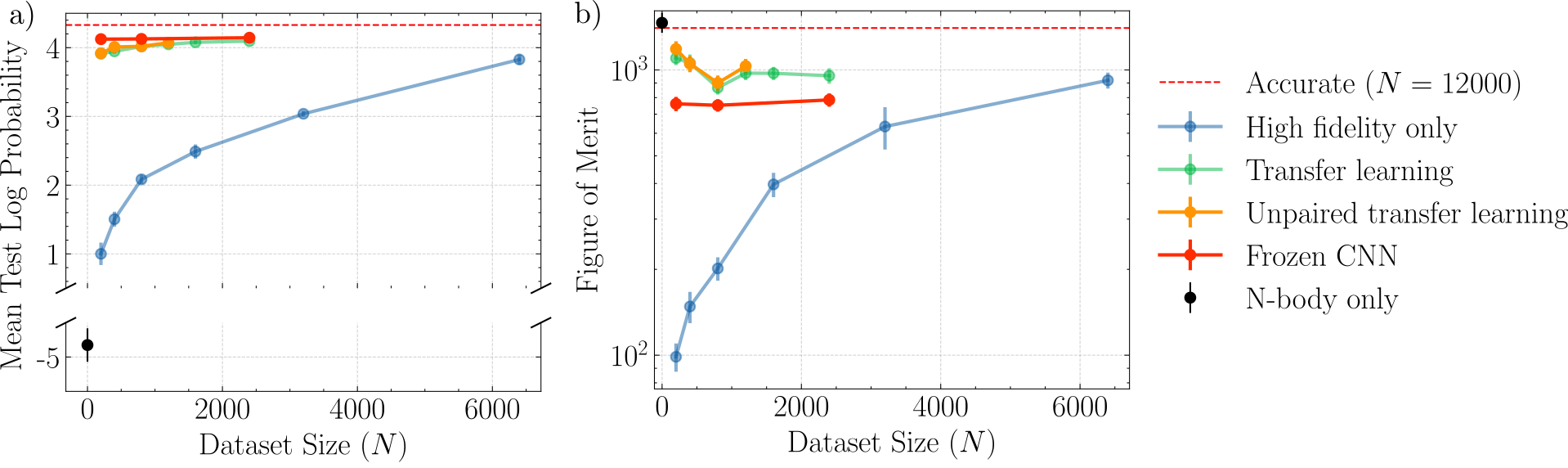}
    \caption{Comparing inference results on the IllustrisTNG LH suite for various experiments. We reproduce the results from \cref{fig:LH_metrics} for a) MTPP (note that the y-axis scale has been shortened for enhanced visualisation) and b) FoM. We also show results from experiments:  pre-training and fine-tuning without any paired data (orange); training a neural compression model with $N$ IllustrisTNG maps and then freezing the CNN compression to train an NDE with the full ($N=12000$) LH training suite (red); and the performance of models only pre-trained on $N$-body simulations (black, corresponding to $N=0$ IllustrisTNG maps).}
    \label{fig:ablations}
\end{figure*}

\begin{figure*}

	\includegraphics[width=\textwidth]{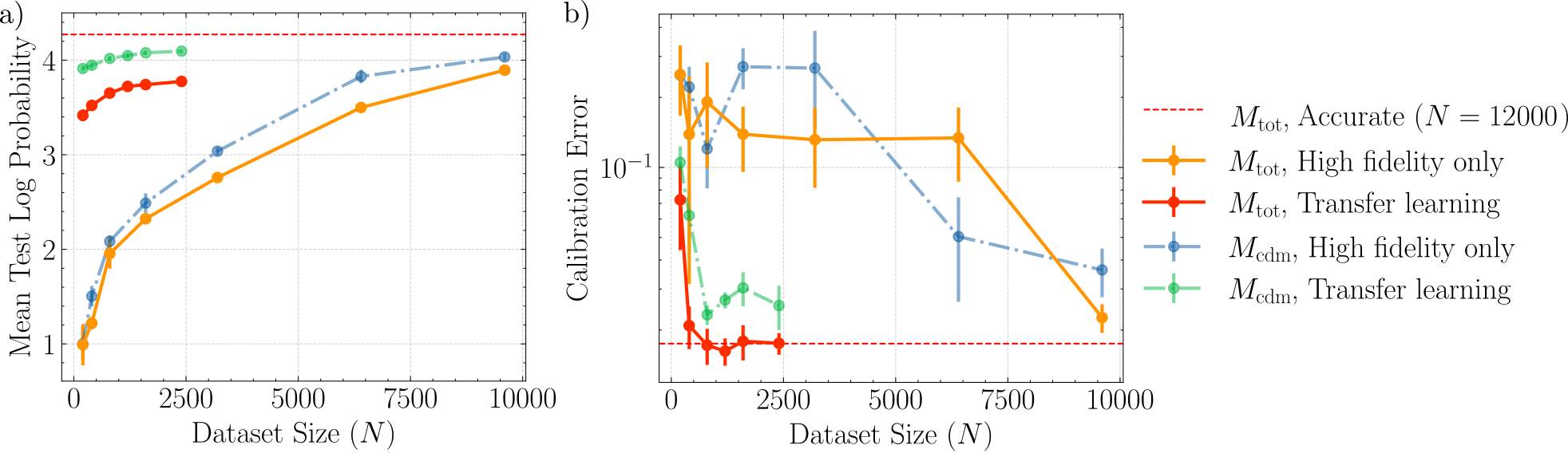}
    \caption{Comparing transfer learning results on the IllustrisTNG LH suite using $M_\textrm{cdm}$ (dot-dashed lines) and $M_\textrm{tot}$ (solid lines). We reproduce the $M_\textrm{cdm}$ results from \cref{fig:LH_metrics} for a) MTPP and b) calibration error. Pre-training on $N$-body simulations gives near equivalent improvements over training with only high-fidelity maps when performing inference on $M_\textrm{tot}$ fields.}
    \label{fig:appendix_mtot}
\end{figure*}

\begin{figure*}

	\includegraphics[width=\textwidth]{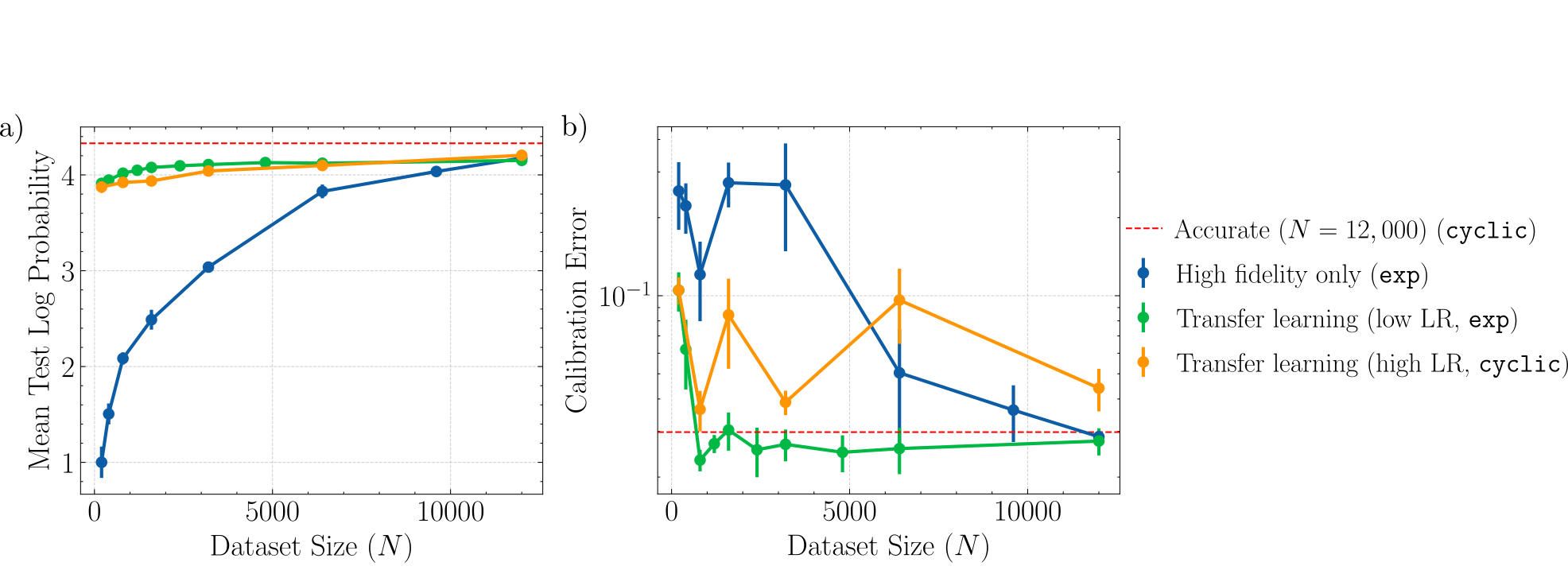}
    \caption{The convergence behaviour of our approaches as a function of dataset size for the LH suite. The legend shows which learning rate scheduler (\texttt{cyclic} or \texttt{exp}) was used, which we found had a minor effect on asymptotic behaviour. We reproduce the $M_\textrm{cdm}$ results from \cref{fig:LH_metrics} and extend the experiments to run all the way to $N=12000$ (blue and green lines). We also report the results of transfer learning using the high learning rate (LR) and cyclic scheduler settings used for the ``Accurate'' baseline model (orange line).}
    \label{fig:appendix_convergence_lr}
\end{figure*}

We performed a range of experiments to better understand the model performance. The results of these experiments are summarised in \cref{fig:ablations}. We explored the impact of the paired aspect of the multifidelity simulation suite (i.e. that each simulation in the lower-fidelity $N$-body suite is paired with a high-fidelity simulation with identical cosmological parameters and initial conditions). This could potentially improve performace, perhaps due to implicit memorisation of the (pre-)training data \citep[a well-studied phenomenon in deep learning, see e.g.][]{yeom2018privacy, carlini2019secret, carlini2022quantifying}. We tested this by ensuring different (i.e. unpaired) cosmologies where used during pre-training and fine-tuning, and found that pairing had no discernable impact on performance. 

We further investigated the neural network inference model by splitting it into two components: the neural compression performed by the CNN, and the density estimation of the NDE. In order to better disambiguate the role of each component, we took the best transfer learning models from \cref{sec:LH_results} and froze the CNN. This fixed the summary statistics that were extracted from the dark matter density maps for a given transfer learning size $N$. We then retrained the NDE using the high-fidelity-only approach on the entire training dataset ($N=12000$) with the frozen neural compression model. The resulting performance is shown in red in \cref{fig:ablations}. The frozen CNN results (red) are produced by taking the CNN from the transfer-learning approach (green), freezing its weights, and retraining the NDE with the full high-fidelity dataset. We observe that at low $N$ (e.g., $N=200$), the model with a frozen CNN but fully retrained NDE (red) outperforms the standard transfer learning baseline (green), suggesting that the NDE limits transfer learning performance in the low $N$ regime. As $N$ increases the baseline transfer learning model catches up, and by $N=2400$ both approaches perform similarly, indicating that the NDE is no longer a performance bottleneck. This convergence implies that the main limitation of transfer learning at higher $N$ (i.e., $N>2400$) is due to slightly suboptimal CNN-based compression. These results suggest that while pre-training on $N$-body simulations encourages highly informative summaries, there are likely subtle differences in the high-fidelity IllustrisTNG simulations (that are useful for slightly improving cosmological constraints) which the CNN fails to discover during fine-tuning.  

\cref{fig:ablations} also shows the $N=0$ transfer learning case, where only the $N$-body simulation pre-training is performed and no high-fidelity maps are used. The extremely poor performance indicates that there are significant differences between the different simulation fidelities, and a fine-tuning step is necessary. 

Next, we test whether transfer learning for (the more observationally important) total matter density $M_\textrm{tot}$ behaves any differently. A key concern is that the improved performance from transfer learning could be largely due to the strong similarity between the IllustrisTNG dark matter density $M_\textrm{cdm}$
and that of dark matter-only $N$-body simulations. \Cref{fig:appendix_mtot} demonstrates that multifidelity transfer learning performs just as well when fine-tuning on $M_\textrm{tot}$.

We compare the results from \cref{sec:LH_results} with inference on the $M_\textrm{tot}$ field with an identical methodology. We find that transfer learning still leads to up to an order-of-magnitude reduction in the number of high-fidelity maps required to train an accurate, trustworthy inference model, compared with high-fidelity-only training. 

The small downward shift of all $M_\textrm{tot}$ inference performance curves (``accurate'', transfer learning and high-fidelity-only) on the MTPP metric from \cref{sec:LH_results}a indicates that inference using the $M_\textrm{tot}$ maps is slightly more challenging. However, there is also a slightly larger gap between the ``accurate'' $M_\textrm{tot}$ model and transfer learned models (and low $N$ high-fidelity-only models) compared with inference results on $M_\textrm{tot}$. This suggests that: i) some features in $M_\textrm{tot}$ require a large number of training maps ($N>6400$) for the CNN to learn to extract, more-so than in the $M_\textrm{cdm}$ case, and ii) $N$-body pre-training gives slightly less informative features than in the  $M_\textrm{cdm}$ case, perhaps for similar reasons as i). 

Finally, we present results exploring the convergence properties of transfer learning and high-fidelity-only training in \cref{fig:appendix_convergence_lr}. We extend the dataset sizes up to $N=12000$ for the high-fidelity-only and transfer learning approaches. The additional line (orange) shows the results from transfer learning when using the same learning rate and scheduler as the ``accurate'' reference models. We find that all models asymptote slightly below the ``accurate'' model baseline. In the high-fidelity-only approach, this is due to a difference in scheduler; as discussed in the text, we find a cyclic learning rate scheduler (``accurate'', red) slightly outperforms an exponential scheduler for large dataset sizes ($N > 10000$). In the transfer learning case, using the same learning rate and scheduler as the ``accurate'' models (orange) gets closer to recovering the baseline at $N=12000$, but this leads to higher variance training with worse calibration at smaller dataset sizes. We therefore conclude that pre-training can lead to a very minor reduction in performance when the pre-training dataset is no larger than the target dataset, but that in all other instances it is desirable.

\label{sec:performance_ablations}

\section{Unconstrained parameters in SB28}
\label{app:SB28}
\begin{figure*}

	\includegraphics[width=0.9\textwidth]{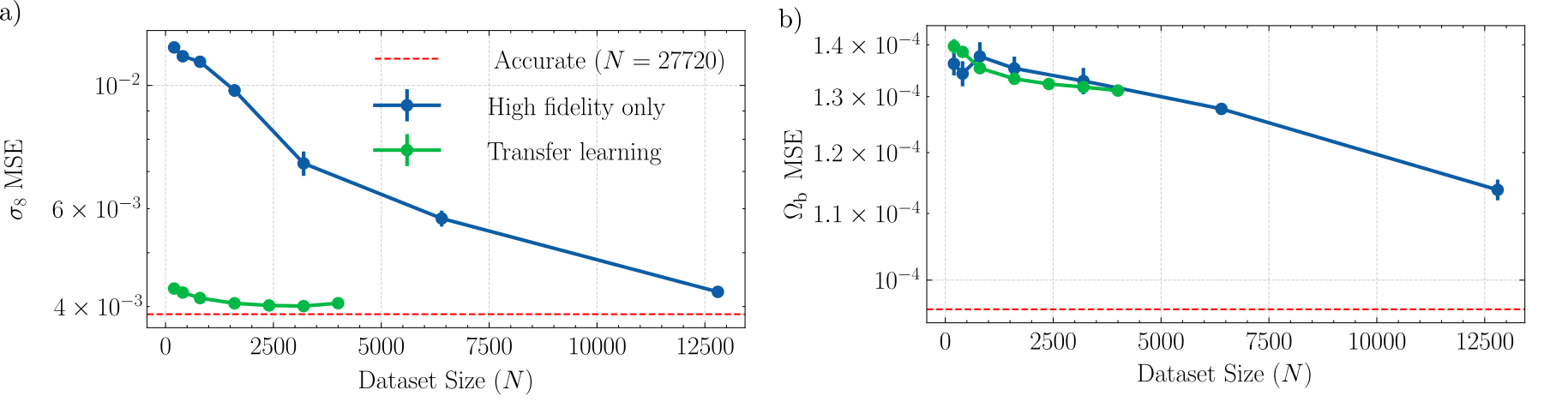}
    \caption{Mean squared error (MSE) between the inferred posterior mean $\hat{\theta}$ and the true cosmology $\theta$. Results are broken down per-parameter, with panel a) showing $\sigma_8$ and panel b) showing $\Omega_\textrm{b}$. While transfer learning yields a significant improvement in $\sigma_8$, we find negligible impact on $\Omega_\textrm{b}$, which has little effect on the $N$-body simulations. }
    \label{fig:SB28_parameter_errors}
\end{figure*}

\begin{figure*}

	\includegraphics[width=\textwidth]{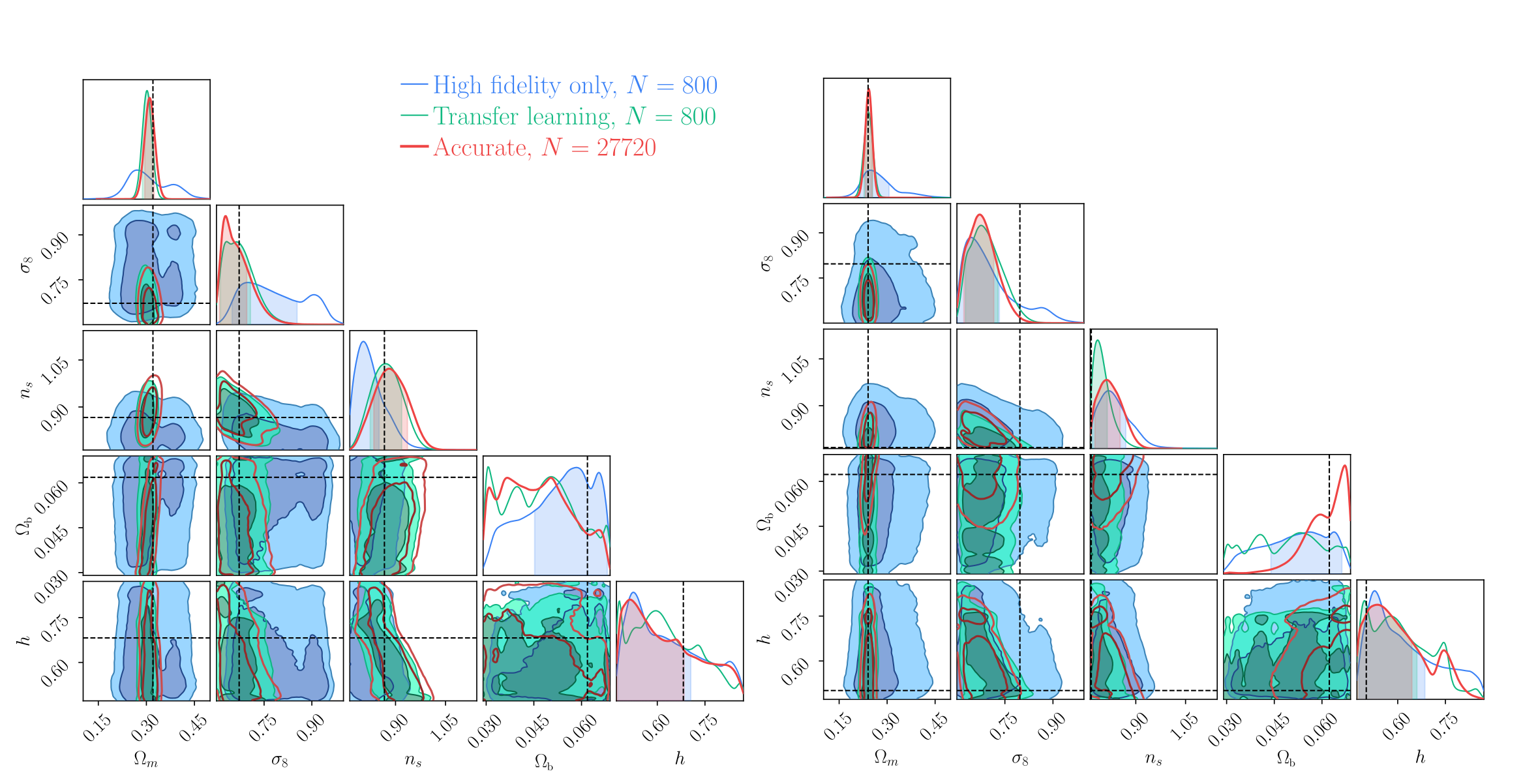}
    \caption{Two examples of posterior inference on IllustrisTNG dark matter maps from the SB28 test suite over the full 5-dimensional posterior. The true cosmology is shown by the black dashed line. A model trained using transfer learning with $N=800$ high-fidelity IllustrisTNG maps is compared against a high-fidelity-only model trained with $N=3200$ maps. The posteriors are compared with an ``accurate'' posterior model that was trained using the full simulation suite. We find that for most 2D dark matter density maps, $\Omega_\textrm{b}$ and $h$ are unconstrained. }
    \label{fig:SB28_posterior_full}
\end{figure*}

The posterior estimation models in \cref{sec:SB28} were trained to perform 5-dimensional inference on $\{\Omega_\textrm{m}, \sigma_8, n_\textrm{s}, h, \Omega_\textrm{b}\}$. However, we found that $h$ and $\Omega_\textrm{b}$ could not be constrained by the data (or, partially, by the CNN-NDE architecture). Here we present some more details on these unconstrained parameters. 

\Cref{fig:SB28_parameter_errors} shows the posterior sample ensemble mean MSE for two cosmological parameters: $\sigma_8$ and $\Omega_\textrm{b}$. The posterior recovery of $\sigma_8$ behaves similarly to \cref{sec:LH_results}, with very good performance relative to the ``accurate'' baseline. On the other hand, we find that pre-training on $N$-body simulations leads to no improvement over the high-fidelity-only training approach for $\Omega_\textrm{b}$. In one sense this is expected: $N$-body simulations do not provide a strong probe of how $\Omega_\textrm{b}$ affects dark matter maps (beyond the initial matter power spectrum), and so there should not be much direct transfer of knowledge. In addition, \cref{fig:SB28_posterior_full} demonstrates that there is little constraining information on $\Omega_\textrm{b}$ in the dark matter density maps.

However, the fact that the high quality pre-trained summary statistics cannot be adapted to improve inference of $\Omega_\textrm{b}$ is, at least naïvely, somewhat surprising. This suggests that the features relevant for inferring $\Omega_\textrm{b}$ are disjoint from those governed by variations in $\{\Omega_\textrm{m}, \sigma_8, n_\textrm{s}, h\}$ for $N$-body simulations, resulting in a representation mismatch that prevents effective transfer from pre-training. \citet{ni2023camels} demonstrated that the values of $\Omega_\textrm{b}$ explored in the CMD SB28 simulation suite had a minor effect on both the star formation rate density and the gas power spectra of the simulations, smaller even than several of the astrophysical nuisance parameters. In addition to this, since  $\Omega_\textrm{b}$ primarily modulates the amount of gas available for star formation and black hole accretion \citep{elbers2025flamingo}, these signatures may be occluded by the wide range of nuisance parameters affecting baryonic feedback in the hydrodynamical simulations.

On the other hand, we found that while $h$ could not be properly constrained by the data, transfer learning yielded similar constraints to the ``accurate'' baseline (as opposed to $\Omega_\mathrm{b}$, which was poorly constrained \textit{and} transfer learning gave no benefit).

Two examples of 5-dimensional posterior inference are given in \cref{fig:SB28_posterior_full}. In the first example, none of the models can constrain $h$ and $\Omega_\textrm{b}$ much beyond the uniform prior. We can highlight two key qualitative features: the fine-tuned approach gives significantly better agreement with the baseline than training from scratch, and it is statistically consistent with the accurate posterior. In the second example, the ``accurate'' posterior gives a (weak) constraint on $\Omega_\textrm{b}$, while the other models fail to provide any constraints. A small but not insignificant fraction of the inferred posteriors follows this second pattern, which is consistent with the minor improvement in constraining power of $\Omega_\textrm{b}$ shown in \cref{fig:SB28_parameter_errors}. We found that these examples tended to coincide with extreme cosmologies (at the boundaries of the prior volume), and particularly for large values of $\Omega_\textrm{b}$ as is the case in \cref{fig:SB28_posterior_full}. We reserve a more systematic analysis as a potential avenue for future work.

\end{document}